\documentclass{aa}

\usepackage{graphicx}
\usepackage{subfigure}
\usepackage{txfonts}
\usepackage{color}
\begin{document}

%\title{Hierarchical fragmentation in the massive star-forming region W51 North}
%\title{W51\,North: a protocluster forming out of a thermally-Jeans fragmenting cloud?}
\title{W51\,North: A protocluster emerging out of a thermally inhibited fragmenting cloud}
\titlerunning{W51\,North: a protocluster emerging out of a thermally inhibited fragmenting cloud}

   \author{Mengyao Tang\inst{1,2,3},
    %\and
    Aina Palau\inst{2},
    Luis A. Zapata\inst{2}, %\fnmsep\thanks{Just to show the usage of the elements in the author field}
    and Sheng-Li Qin\inst{1}
          }

   \institute{
   Department of Astronomy, Yunnan University, and Key Laboratory of Astroparticle Physics of Yunnan Province, Kunming, 650091, People's Republic of China\\
  %\email{mengyao\_tang@yeah.net}
  \and
  Instituto de Radioastronom\'ia y Astrof\'isica, Universidad Nacional Aut\'onoma de M\'exico, P.O. Box 3-72, 58090, Morelia, Michoac\'an, M\'exico \\
    \and
    Institute of Astrophysics, School of Physics and Electronical Science, Chuxiong Normal University, Chuxiong 675000, China \\
   \email{l.zapata@irya.unam.mx, a.palau@irya.unam.mx}
   %\thanks{The university of heaven temporarily does not accept e-mails}
             }

   %\date{Received September 15, 1996; accepted March 16, 1997}

% \abstract{}{}{}{}{}
% 5 {} token are mandatory
\authorrunning{Mengyao Tang et al}

  \abstract
{The fragmentation process in massive star-forming regions is one of the contemporary problems in astrophysics,
    and  several physical processes have been proposed to control the fragmentation including turbulence,
    magnetic field, rotation, stellar feedback, and gravity. However, the fragmentation process has been poorly studied at small spatial scales well below 1000 AU.}
   {W51\,IRS2 is a well-known massive star-forming region whose fragmentation properties have not been thoroughly investigated yet.
        We aim to use the Atacama Large Millimeter and Submillimeter Array (ALMA) high angular resolution data to identify the fragments in W51\,IRS2 and to study the
        fragmentation properties on a spatial scale of 0.04$^{\prime\prime}$ (200 AU).}
   {We used ALMA data of W51\,IRS2 from three projects, which give an angular resolution of 0.028$^{\prime\prime}$ (144 AU) at millimeter wavelengths. The continuum images reveal a significant substructure in an east-west ridge, where both W51\,North and W51d2 are embedded.
        A spectral index map has been generated from the 3 and 1.3 mm high-resolution continuum images.
        We identified compact fragments by using {\it uv}-range constrained 1.3 mm continuum data.
   A mean surface density of companions (MSDC) analysis has been performed to study the separations between fragments.}
   {A total number of 33 continuum sources are identified and 29 out of them are defined as fragments in the surveyed region.The MSDC analysis reveals two breaks corresponding to spatial scales of 1845 AU and 7346 AU, indicative of a two-level clustering phenomenon, along with a linear regime below 1845 AU, mostly associated with W51 North, whose slope is consistent with the slope for the clustering regime of other cluster-like regions in the Galaxy.}
        {The typical masses and separations of the fragments as well as the relation between the density and number of fragments can be explained through a thermal Jeans process operating at high temperatures of 200--400 K, consistent with previous measurements of the temperature in the region, and produced by the nearby massive stars. Therefore, although W51\,IRS2 seems to be undergoing a thermally inhibited fragmentation phase, this does not seem to prevent the formation of a protocluster associated with W51\,North.
        }


   \maketitle

%
%-------------------------------------------------------------------

\section{Introduction}
\label{introduction}
In the past few decades, several studies have been performed to investigate the properties of fragmentation in molecular clouds.
These works have concentrated on the study of the physical processes that control fragmentation such as turbulence
\citep{Zhang2009,Zhang2015,Pillai2011,Wang2011, Wang2014,Lu2015}, disk fragmentation \citep[e.g.,][]{Adams1989,Kratter2010,Vorobyov2010},
rotational fragmentation \citep[e.g.,][]{Lim2016,Arreaga-Garcia2017}, stellar feedback \citep[e.g.,][]{Krumholz2007,Peters2010,Myers2013}, and magnetic fields
 \citep[e.g.,][]{Boss2004,Vazquez-Semadeni2005,Commercon2011a,Hennebelle2011}.
On the other hand, a number of studies have found that the fragmentation of dusty cores seems to be related to the density of the cores \citep[e.g.,][]{Gutermuth2011,Lombardi2013,Ragan2013,Imara2015,Lee2015,Mairs2016,Nguyen-Luong2016,Liu2016,Pokhrel2016,Pokhrel2018,Hacar2017,Mercimek2017,Alfaro2018,Li2019,Orkisz2019,Sanhueza2019,Sokol2019,Svoboda2019,Zhang2019},
indicating that the fragmentation of cores is consistent with a thermal Jeans process in these cases.

Among the aforementioned processes potentially controlling fragmentation, both the magnetic field and stellar feedback have been poorly studied so far from an observational point of view. 
While for the magnetic field there are very recent works aimed at studying its effects in samples of massive dense cores where both fragmentation and magnetic field properties have been observed \citep[e.g.,][]{AnezLopez2020,Palau2021}, the stellar feedback effects have not been broadly studied yet.

In addition, the concept of hierarchical fragmentation \citep{Hoyle1953} has been explored in several recent observational studies toward low- and high-mass star-forming regions \citep{Teixeira2006,Teixeira2007,Zhang2009,Wang2011,Hacar2013,Pineda2013,Wang2014,Teixeira2016,Kainulainen2017,Pokhrel2018},
 suggesting that hierarchical fragmentation is a common process \citep{Wang2011, Zhang2015,Teixeira2016,Kainulainen2017}.
Therefore, it is very important to study the processes that control fragmentation in more star-forming regions, paying special attention to the massive ones, which are typically associated with complex structures and feedback processes. However, given the large distances of massive star-forming regions, the spatial scales of 200 AU ($\sim$ 0.001 pc) have been poorly explored to date.

The W51 IRS2 region \citep{Ginsburg2017} harbors a well-known H$_{\rm II}$
region with a luminosity of $\sim$ 3$\times$10$^{6}$ $L_{\sun}$ \citep{Erickson1980}, and it is located at 5128($\pm$1867) pc \citep{Reid2019}.
In the W51 IRS2 region, there are two highly obscured massive sources called W51 North and W51 d2.
W51 d2 was reported as an ultracompact H$_{\rm II}$ (UCH$_{\rm II}$) region \citep{Gaume1993,Lacy2007},
and  W51 North, located to the east of W51 d2, is an active massive star-forming region \citep{Zhang1998,Sollins2004,Zapata2008,Zapata2009,Zapata2010}.
W51 North shows strong emission at millimeter and submillimeter wavelengths, very faint centimeter free-free emission \citep{Gaume1993,Zhang1998,Eisner2002,Zapata2009},
and no mid-infrared counterparts \citep{Kraemer2001,Okamoto2001,Lacy2007}.
In addition, very strong OH, SiO, and H$_{2}$O maser emissions have been detected \citep{Schneps1981,Gaume1987,Morita1992}.
This emission is associated with the formation of high-mass stars.
Fortunately, W51\,IRS2 was already been observed by the Atacama Large Millimeter and Submiilimeter Array (ALMA) at different wavelengths with several angular resolutions
(the best angular resolution of 0.028$^{\prime\prime}$ corresponds to 144 AU).
The high angular resolution capability of ALMA gives us a chance to reveal the small-scale properties of this molecular cloud.

The paper has been organized as follows.
Section~\ref{observation} introduces the archival data and the reduction of all observations. Section~\ref{results}
presents our results directly derived from the observations. Section~\ref{analyses} and Section~\ref{discussion} provide detailed analyses
and a discussion on the fragmentation in W51\,IRS2. The main findings are summarized in Section~\ref{conclusions}.

\section{Observational data}
\label{observation}

In this work, we used three ALMA datasets of W51 IRS2 from the following projects: 2017.1.00293.S, 2015.1.01596.S, and  2013.1.00994.S.
Using the pipeline in CASA \citep{McMullin2007} (Version 4.7), we obtained the visibilities to perform self-calibration and generate the final continuum images.
The continuum images of calibrated visibilities were produced with the task TCLEAN of the CASA package.
During the imaging process, for all datasets, a ``briggs'' weighting ROBUST parameter of 0.5 was set to balance the sensitivity and angular resolution.

\subsection{ALMA 3 mm data}
\label{3 mm}

Band 3 (3 mm) observations were carried out with ALMA from 2017 October 6 to 15 as part of Cycle  5 program 2017.1.00293.S. The project was carried out in
five executions using between 48 and 51 antennas with a diameter of 12 m. The baselines range from 33 to 15983 m (9.389 -- 5.33$\times$10$^{3}$ k$\lambda$),
which corresponds to the largest recoverable angular scale of 9.7$^{\prime\prime}$ (following equation (A.5) of \citet{Palau2010}). 
The observing field of W51 IRS2  was centered
at the following sky position: RA(J2000) = 19$^{h}$23$^{m}$40$^{s}$, Decl (J2000) = +14$^{\circ}$31$^{\prime}$05$^{\prime\prime}$.
Strong maser lines were included in the 3 mm data \citep{Ginsburg2019}. 
Additionally, W51 North is a well-known chemically rich hot core. 
In order to avoid line contamination, the 3 mm continuum image was only generated using the line-free channels.
The channel selection was performed in the {\it uv} domain, and the line-free channels were discerned in the four spectral windows, which have a bandwidth of 1.875 GHz, and they are centered at rest frequencies of 86.919 GHz (spw0), 87.828 GHz
(spw1), 98.229 GHz (spw2), and 99.978 GHz (spw3). The quasar J2148$+$0657 was adopted as a flux and bandpass calibrator, and the phase gain calibrator was J1922$+$1530.

We performed six rounds of phase self-calibration and one round of amplitude self-calibration to visibilities.
The phase-only self-calibrations decreased the root mean square (rms) noise of continuum image from 0.5 mJy beam$^{-1}$ to 0.15 mJy beam$^{-1}$.
The amplitude self-calibration did not introduce any systematic error, yielding a final continuum image with a rms noise of 0.08 mJy beam$^{-1}$ and a synthesized beam of 0.078$^{\prime\prime}$$\times$0.051$^{\prime\prime}$ (position angle: --45$^{\circ}$).

\subsection{ALMA 1.3 mm data}
\subsubsection{High angular resolution}
\label{1.3 high}
Band 6 (1.3 mm) high angular resolution observations were carried out with ALMA in 2015 October 30 and 31 as part of Cycle 3 program 2015.1.01596.S.
The project was performed in two executions using 42 antennas with a diameter of 12 m. The baselines range from 40 to 16161 m (51.563 -- 1.27$\times$10$^{4}$ k$\lambda$),
corresponding to the largest recoverable angular scale of 1.6$^{\prime\prime}$. The phase center of the observations is at the sky position RA(J2000) = 19$^{h}$23$^{m}$40$^{s}$.05, Decl (J2000) = +14$^{\circ}$31$^{\prime}$05$^{\prime\prime}$.50.
The continuum image is obtained by averaging the line-free channels from ten spectral windows.
There are two spectral windows with a 1.875 GHz bandwidth,  and the remaining eight spectral windows have a 234 MHz bandwidth.
The rest frequencies are 217.840 GHz (spw0), 221.310 GHz (spw1), 220.650 GHz (spw2), 219.560 GHz (spw3),
220.400 GHz (spw4), 233.296 GHz (spw5), 232,164 GHz (spw6), 232.861 GHz (spw7), 231.901 GHz (spw8), and 235 GHz (spw9).
The flux, bandpass, and phase calibrators were J2148+0657, J1751+0939, and J1920+1540, respectively.

We performed five rounds of phase self-calibration and one round of amplitude self-calibration to visibilities.
The phase-only self-calibrations decreased the rms noise of the 1.3 mm high-resolution continuum image from 0.2 mJy beam$^{-1}$ to 0.1 mJy beam$^{-1}$, while an additional 0.05 mJy beam$^{-1}$ systematic error from amplitude self-calibration was introduced to the 1.3 mm high-resolution continuum data,
yielding a final continuum image with a rms noise of 0.15 mJy beam$^{-1}$ and a synthesized beam of 0.031$^{\prime\prime}$$\times$0.025$^{\prime\prime}$ (position angle: --45$^{\circ}$).

\subsubsection{Low angular resolution}
\label{1.3 low}
Band 6 (1.3 mm) low angular resolution observations were carried out with ALMA in 2013 July 18 as part of Cycle 2 program 2013.1.00994.S.
The project was carried out using 38 antennas with a diameter of 12 m. 
The baselines range from 124 to 1573 m (9.164 -- 1573.35 k$\lambda$), corresponding to the largest recoverable angular scale of 9.9$^{\prime\prime}$. The phase center of the observations is at the sky position RA(J2000) = 19$^{h}$23$^{m}$39$^{s}$.95,
Decl (J2000) = +14$^{\circ}$31$^{\prime}$05$^{\prime\prime}$.50.
The continuum image was obtained by averaging all channels from four spectral windows, which have a bandwidth of 1.875 GHz and they are centered
at rest frequencies of 224.984 GHz (spw0), 226.984 GHz (spw1), 239.015 GHz (spw2), and 241.015 GHz (spw3). The quasars J1924$-$2194 and J1751$+$0939 were used as bandpass calibrators. 
The flux and phase calibrators are Titan and J1922+1530, respectively.
The aim of this project is to search for polarization in W51 IRS2, consequently the polarization calibrator was J1924-2914, but we only took data from the Stokes $^{\prime} I {^\prime}$ parameter to generate the continuum image.

We performed seven rounds of phase self-calibration and one round of amplitude self-calibration to the visibilities.
The phase-only self-calibrations decreased the rms noise of the 1.3 mm low-resolution continuum image from 1.5 mJy beam$^{-1}$ to 0.7 mJy beam$^{-1}$. 
The amplitude self-calibration did not introduce any systematic errors,
yielding a final continuum image with a rms noise of 0.5 mJy beam$^{-1}$ and a synthesized beam of 0.237$^{\prime\prime}$$\times$0.217$^{\prime\prime}$ (position angle: 28$^{\circ}$). 

\label{continuum}
\begin{figure*}[!h]
        \centering
        \includegraphics[width=\hsize]{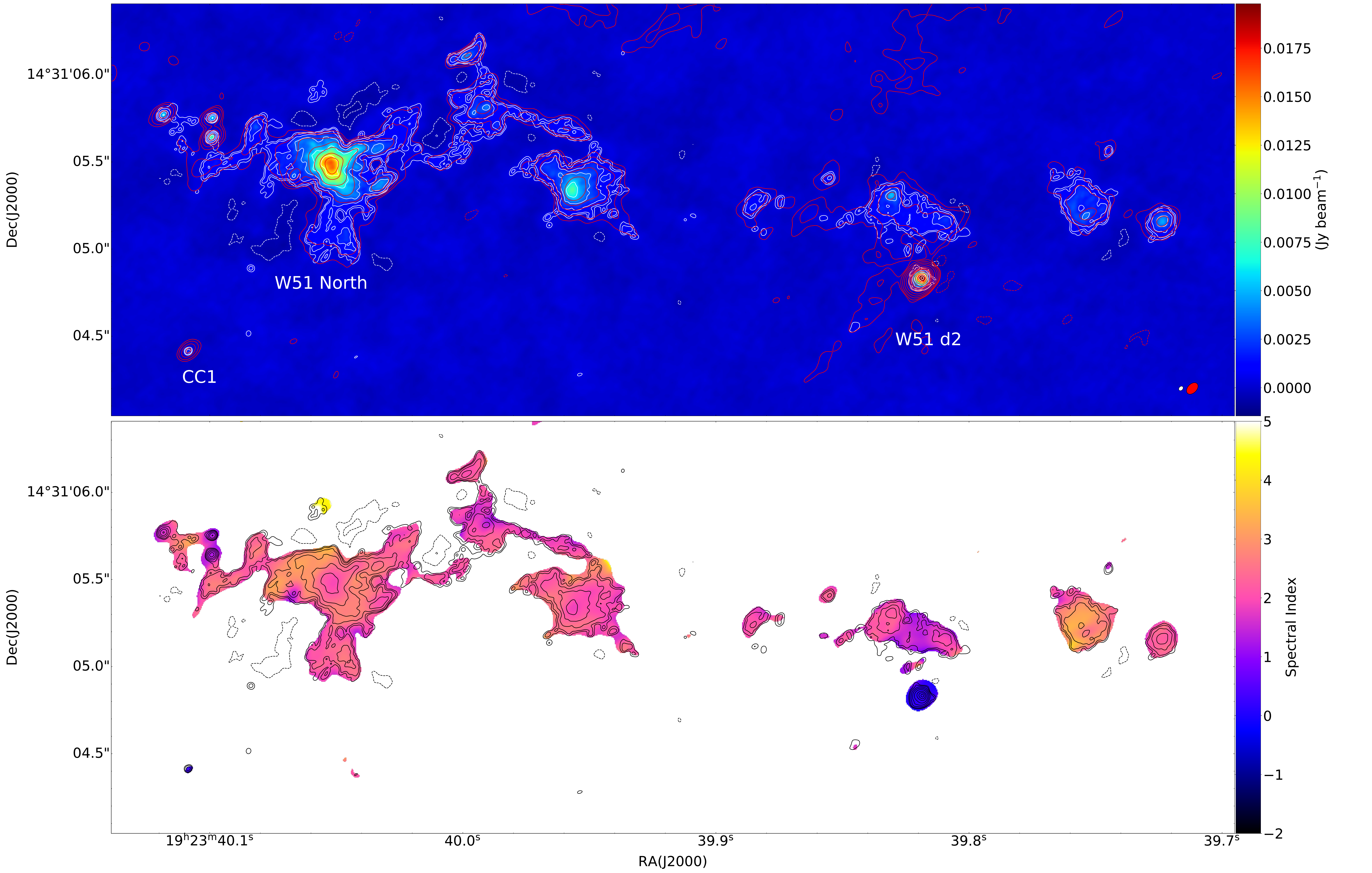}
        \caption{ \emph{Upper panel}: Color image and white contours represent the 1.3 mm high-resolution continuum map
        of the W51\,IRS2 region. Contour levels are [-5,-3, 3, 5, 10, 20, 40, 80, 120, 160]$\times\sigma$ ($\sigma$=0.15 mJy beam$^{-1}$).
        Red contours correspond to the 3 mm continuum map, and they are [-3, 3, 5, 10, 20, 40, 80, 120, 160, 200, 300, 400, 500]$\times\sigma$ ($\sigma$=0.08 mJy $^{-1}$).
        The red and white ellipses in the bottom-right corner indicate the synthesized beams of the 3 and 1.3 mm high-resolution continuum maps, respectively.
        \emph{Bottom panel}: Color image represents the spectral index map that was derived from the 1.3 and 3 mm high-resolution continuum maps.
        The black contours are the 1.3 mm high-resolution continuum, and the contour levels are the same as in the upper panel.}
        \label{continuum_fig}
\end{figure*}

\section{Results}
\label{results}

\subsection{Dust emission and spectral index}
\label{spectral index}

In the upper panel of Fig.~\ref{continuum_fig}, we present the 3 and 1.3 mm ALMA high-resolution continuum images obtained in this study.
The color scale and white contours correspond to the intensity distribution of the 1.3 mm high-resolution continuum, and the red contours correspond to the intensity
distribution of the 3 mm continuum, respectively.
The sources corresponding to W51 North and W51 d2 are labeled in this figure.
W51 IRS2 was previously observed by several projects, where four clumps (SMA1--4) were reported in this region following a filamentary structure
along the east-west direction \citep{Zapata2009,Tang2013, Koch2018}.
The upper panel of Fig.~\ref{continuum_fig} shows that most of the continuum objects are well resolved, and the majority of these sources have 3 mm counterparts.
Most of the detected objects are part of elongated structures, consistent with previous observations. 
Our data confirm that W51\,IRS2 is a complex and structured region.

The bottom panel of Fig.~\ref{continuum_fig} presents a spectral index map that was derived from the 3 and 1.3 mm ALMA high angular resolution continuum images.
The 1.3 mm high-resolution continuum data are overlaid in the spectral index map as black contours to indicate the structure of the continuum emission.
The spectral index map was derived from the ``spix'' mode in the ``immath'' task of the CASA package \citep{McMullin2007}, which computes the spectral index using two continuum images with different frequencies based on the relationship between flux density and spectral index of $S_{\nu}\propto\nu^{\alpha}$.
The spectral index is defined as  $\alpha$=$ln(I_{3mm}/I_{1.3mm})/ln(\nu_{3mm}/\nu_{1.3mm})$, where $I$ is the intensity value of each pixel and $\nu$ is the frequency.
Since these two continuum images have different synthesized beams, before generating the spectral index map,
we tapered them to the same resolution of 0.079$^{\prime\prime}\times$0.053$^{\prime\prime}$ using a pixel size of 0.01$^{\prime\prime}$.
In order to produce this spectral index map, we also included a cut-off with rms noises higher than 3$\sigma$. 
This helped us discriminate the fake sources in the field.
Based on the spectral index map, most sources have spectral indices of around 2, suggesting that the dust emission may be partly optically thick since optically thin dust emission is expected to have a substantially steeper spectral index in the range of 3--4.
%, indicating that the majority of continuum objects are associated with thermal dust emission.
The averaged spectral index over the W51 North region is $\sim$2.5, which is comparable to the previously reported spectral index of 2.8 in W51 North using Submillimeter Array  observations \citep{Zapata2009}.
W51 d2 and the region that we call CC1, which are two small regions located to the south of the surveyed area, show spectral indices around 0.2, which is related to optically thin free-free emission from a H$_{\rm II}$ region.  This is also consistent with previous works finding that W51 d2 is a very
bright centimeter continuum source \citep{Zapata2009,Ginsburg2016,Ginsburg2017}.

\begin{table*}
        \caption{\label{frag_table1}Continuum sources identified in the $uv$-range constrained 1.3 mm continuum image}
        \centering
        \tiny
        \begin{tabular}{lcccccccccc}
                \hline\hline
                Name &RA(J2000)&Dec(J2000)  &\multicolumn{2}{c}{Source size}     &P.A     &Intensity   &Flux density  &Mass  &Mass$_{\rm t}^{\ast}$ &SPIX$^{\star}$\\
                \cline{4-5}
                &                       &                              &major($^{\prime\prime}$)  &minor($^{\prime\prime}$)        &($^{\circ}$)    &(mJy beam$^{-1}$)  &(mJy) &($M_{\sun}$)  &($M_{\sun}$) &\\
                \hline
CA1     $\dagger$  &19:23:40.11825  &+14:31:05.7697  &0.016(0.001)      &0.011(0.001)   &122.0  &5.52(0.09)  &7.0(0.2)   &0.3(0.2)    &0.04(0.01) &1.65(0.04)\\
CA2     $\dagger$  &19:23:40.09887  &+14:31:05.7533  &0.013(0.001)      &0.007(0.001)   &137.0  &6.09(0.07)  &7.0(0.1)   &0.3(0.2)    &0.02(0.01) &1.32(0.06)\\
CA3     $\dagger$  &19:23:40.09924  &+14:31:05.6428  &0.018(0.002)      &0.014(0.001)   &149.0  &8.1(0.2)    &11.4(0.4)  &0.5(0.4)    &0.07(0.01) &1.34(0.04)\\
CA4                &19:23:40.08164  &+14:31:05.7044  &0.037(0.004)      &0.033(0.004)   &110.0  &1.8(0.1)    &5.4(0.4)   &0.2(0.2)    &0.3(0.1)    &2.44(0.04)\\
CB1                &19:23:40.06676  &+14:31:05.4183  &0.026(0.004)      &0.016(0.005)   &168.0  &2.6(0.1)    &4.3(0.3)   &0.2(0.2)    &0.10(0.04) &1.47(0.05)\\
CB2                &19:23:40.05072  &+14:31:05.5984  &0.050(0.001)      &0.017(0.001)   &156.9  &1.66(0.03)  &4.2(0.1)   &0.2(0.1)    &0.22(0.02) &2.65(0.03)\\
CB3            &19:23:40.05154  &+14:31:05.5644  &0.069(0.001)  &0.022(0.000)   &154.0  &1.76(0.02)  &6.50(0.09) &0.3(0.2)    &0.39(0.01) &2.39(0.05)\\
CB4            &19:23:40.05471  &+14:31:05.4955  &0.086(0.001)  &0.055(0.001)   &148.0  &3.76(0.05)  &32.1(0.5)  &1.5(1.1)    &1.24(0.04) &2.19(0.04)\\
CB5                &19:23:40.05197  &+14:31:05.5001  &0.102(0.002)      &0.055(0.001)   &5.5    &4.31(0.07)  &43.4(0.7)  &2.0(1.5)    &1.45(0.05) &2.02(0.04)\\
CB6                &19:23:40.05131  &+14:31:05.4715  &0.077(0.002)      &0.053(0.001)   &166.0  &4.52(0.09)  &33.7(0.8)  &1.5(1.2)    &1.05(0.05) &1.98(0.03)\\
CB7                &19:23:40.05224  &+14:31:05.4304  &0.131(0.002)      &0.056(0.001)   &86.0   &3.41(0.04)  &44.4(0.6)  &2.0(1.5)    &1.91(0.05) &2.13(0.04)\\
CB8                &19:23:40.04953  &+14:31:05.4307  &0.130(0.005)      &0.043(0.002)   &47.1   &3.2(0.1)    &34.4(1.4)  &1.6(1.2)    &1.45(0.13) &2.21(0.07)\\
CB9                &19:23:40.05275  &+14:31:05.3931  &0.063(0.002)      &0.047(0.002)   &165.0  &2.79(0.07)  &15.8(0.5)  &0.7(0.5)    &0.77(0.05) &2.35(0.08)\\
CB10           &19:23:40.04627  &+14:31:05.4067  &0.045(0.002)  &0.021(0.001)   &74.2   &1.63(0.04)  &4.5(0.1)   &0.2(0.2)    &0.25(0.02) &2.5(0.06)\\
CB11           &19:23:40.04359  &+14:31:05.3956  &0.058(0.002)  &0.044(0.000)   &145.0  &1.37(0.03)  &3.4(0.1)   &0.2(0.1)    &0.66(0.02) &2.70(0.06)\\
CB12           &19:23:40.04434  &+14:31:05.3222  &0.046(0.001)  &0.032(0.001)   &46.4   &1.39(0.03)  &4.8(0.1)   &0.2(0.2)    &0.38(0.03) &2.72(0.06)\\
CB13           &19:23:40.04525  &+14:31:05.5315  &0.068(0.002)  &0.039(0.001)   &139.0  &2.90(0.08)  &15.2(0.5)  &0.7(0.5)    &0.69(0.05) &2.43(0.09)\\
CB14           &19:23:40.04219  &+14:31:05.5689  &0.090(0.003)  &0.046(0.002)   &115.0  &3.8(0.1)    &29(1)      &1.3(0.9)    &1.07(0.08) &2.4(0.1)\\
CB15           &19:23:40.03312  &+14:31:05.5822  &0.064(0.006)  &0.046(0.005)   &111.0  &2.2(0.2)    &12(1)      &0.6(0.5)    &0.8(0.2)   &2.54(0.05)\\
CB16           &19:23:40.03449  &+14:31:05.5663  &0.049(0.005)  &0.041(0.005)   &116.0  &2.0(0.1)    &8.2(0.7)   &0.4(0.3)    &0.5(0.1)   &2.63(0.03)\\
CB17           &19:23:40.04159  &+14:31:05.4697  &0.044(0.006)  &0.022(0.005)   &96.3   &2.1(0.2)    &5.6(0.6)   &0.3(0.2)    &0.25(0.08) &2.81(0.04)\\
CB18           &19:23:40.03186  &+14:31:05.3790  &0.093(0.003)  &0.045(0.002)   &99.0   &2.75(0.07)  &21.4(0.6)  &1.0(0.7)    &1.08(0.07) &2.02(0.03)\\
CB19           &19:23:40.03367  &+14:31:05.3562  &0.085(0.002)  &0.042(0.001)   &106.0  &2.87(0.06)  &19.4(0.4)  &0.9(0.7)    &0.92(0.05) &2.29(0.1)\\
CC1$\dagger$   &19:23:40.10853  &+14:31:04.4129  &0.011(0.004)  &0.009(0.004)   &153.0  &2.24(0.08)  &2.6(0.2)   &0.12(0.09)  &0.03(0.02) &0.27(0.03)\\
CD1                &19:23:40.00351  &+14:31:06.0858  &0.058(0.003)      &0.032(0.002)   &122.0  &1.54(0.06)  &6.1(0.3)   &0.3(0.2)    &0.48(0.05) &2.10(0.03)\\
CD2                &19:23:39.99919  &+14:31:06.1008  &0.070(0.005)      &0.030(0.003)   &115.3  &2.9(0.2)    &13.0(0.9)  &0.6(0.5)    &0.54(0.09) &2.07(0.04)\\
CD3                &19:23:39.99575  &+14:31:05.7784  &0.033(0.002)      &0.020(0.002)   &28.8   &1.62(0.05)  &3.5(0.1)   &0.2(0.1)    &0.17(0.03) &2.05(0.01)\\
CD4                &19:23:39.99034  &+14:31:05.8124  &0.034(0.002)      &0.030(0.002)   &133.0  &1.90(0.07)  &4.9(0.2)   &0.2(0.2)    &0.26(0.03) &1.50(0.06)\\
CE1                &19:23:39.95698  &+14:31:05.3403  &0.087(0.007)      &0.056(0.005)   &7.8    &2.8(0.2)    &25(2)      &1.1(0.9)    &1.3(0.2)   &1.98(0.08)\\
CF1                &19:23:39.83064  &+14:31:05.3112  &0.039(0.005)      &0.039(0.005)   &92.0   &2.8(0.2)    &9.4(0.7)   &0.4(0.3)    &0.4(0.1)   &2.18(0.03)\\
CF2                &19:23:39.81841  &+14:31:04.8348  &0.041(0.003)      &0.033(0.003)   &47.0   &18.7(0.7)   &60(3)      &3(2)        &0.35(0.05) &0.16(0.04)\\
CG1                &19:23:39.75344  &+14:31:05.1947  &0.029(0.001)      &0.016(0.000)   &139.0  &1.47(0.01)  &2.57(0.03) &0.12(0.09)  &0.12(0.00) &2.59(0.02)\\
CG2                &19:23:39.72336  &+14:31:05.1534  &0.067(0.005)      &0.057(0.005)   &101.0  &2.3(0.1)    &16(1)      &0.7(0.6)    &1.0(0.2)   &2.25(0.03)\\
                \hline
        \end{tabular}
        \tablefoot{The flux of each source was derived from the 1.3 mm continuum image with a {\it uv} range > 910 k$\lambda$. The errors of the source size, flux density, and intensity mainly come from the two-dimensional Gaussian fitting. The error in mass is mainly due to uncertainties as to the flux density and distance, as well as temperature and dust opacity; however, here, only the flux density and distance were taken into account.\\
        $\dagger$: Sources marked with ``$\dagger$'' are isolated and point-like sources. They are not defined as fragments and are excluded from further analysis. \\
        $\ast$: Masses estimated assuming the continuum sources are optically thick, adopting a column density of $N_{\rm H_{2}}$ = $\frac{1}{\kappa_{\nu}}$ = 2.6$\times$10$^{25}$ cm$^{-2}$ for the 1.3 mm data. \\
    $\star$:The averaged spectral index of each fragment was measured from the spectral index map. The error is the standard deviation.}
\end{table*}

\subsection{Fragment identification}
\label{fragments}
%As we mentioned in the previous Section, a large amount of continuum objects have been detected, and the W51 IRS2 region seems to be undergoing fragmentation.
To study the fragmentation properties, it is necessary to identify fragments in this region.
The 1.3 mm high-resolution image is the best option to do this job as it has the highest angular resolution of 0.028$^{\prime\prime}$
(corresponding to a spatial scale of 144 AU at 5128 pc distance).
As we mention in Section~\ref{1.3 high}, the largest recoverable angular scale for this dataset is 1.6$^{\prime\prime}$.
This means that the flux from the 1.3 mm high-resolution continuum may not only come from compact structures, but it also may come from their extended surrounding environment.
As shown in the upper panel of Fig. \ref{continuum_fig}, the majority of the 1.3 mm continuum objects from the continuum image seem to have angular sizes smaller than 0.2$^{\prime\prime}$ ($\sim$1000 AU).
Therefore, to identify the real compact and small-scale fragments, we constrained  the 1.3 mm high-resolution continuum image by using only
visibilities with a {\it uv} range larger than 910 k$\lambda$ (corresponding to the largest recoverable angular scale of 0.1$^{\prime\prime}$).
However, cutting off short baselines may lead to missing flux problems and create artifacts in the continuum image. 
In order to test whether the short baselines removed here can significantly affect our results, we simulated the ALMA observations using the ``SIMOBSERVE'' task in the CASA package.
The detailed description and results of the simulations are presented in Appendix \ref{AppendixA}. 
Based on the results of our simulations, it seems that cutting off baselines shorter than 910 k$\lambda$ does not create significant artifacts, but it allowed us to recover the real compact sources.

\begin{figure*}[]
        \centering
        \includegraphics[width=\hsize]{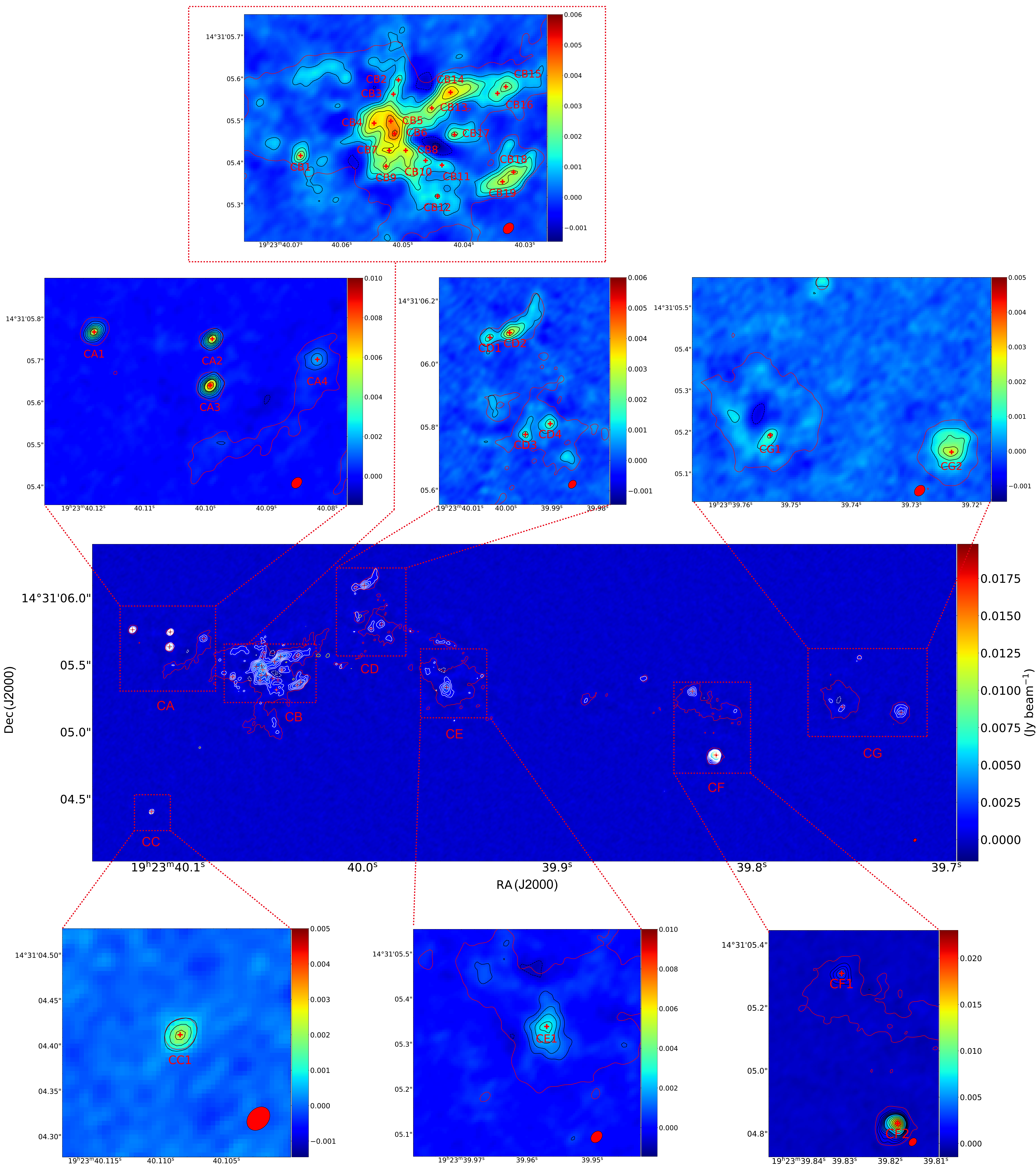}
        \caption{Color image and white contours correspond to the {\it uv}-range constrained 1.3 mm high-resolution continuum (with a {\it uv} range larger than 910 k$\lambda$).
                The white contour levels are [-12, -6, 6, 12, 18, 24, 30, 36, 40, 60, 80, 100, 120, 140, 160]$\times\sigma$ ($\sigma$ = 0.11 mJy beam$^{-1}$),
                and the beam size is shown in the bottom-right corner of each plot. The red contour in each plot is the full {\it uv}-range 1.3 mm high-resolution continuum image,
                and the contour level is 6$\sigma$ ($\sigma$ = 0.15 mJ beam$^{-1}$). 
       We identified fragments as sources with at least one closed contour above the 12$\sigma$ (1.32 mJy beam$^{-1}$) identification threshold.
                A total of 33 continuum objects have been identified in the region.
                To illustrate the ``closed contour'' identification method, we plotted the first closed contour above the threshold that separates each identified source as an orange contour in Fig. \ref{dendrogram}.
                To give a better view of the fragments, we divided the whole region into seven subclumps (clump A to G are named CA, CB, CC, CD, CE, CF, and CG), shown in the individual close-in panels, where contours are shown in black to improve the contrast over the background.
                The crosses in the subplots indicate the central positions of the identified fragments. The names of the fragments are labeled.
                A Gaussian fitting was performed to obtain the basic parameters of the identified objects and the fitting results are listed in Table~\ref{frag_table1}. }
        \label{1.3mm_910klambda}
\end{figure*}

Fig.~\ref{1.3mm_910klambda} shows the 1.3 mm continuum image obtained from the data with a {\it uv} range larger than 910 k$\lambda$.
This restriction on the {\it uv} range results in a rms noise of $\sigma$ = 0.11 mJy beam$^{-1}$.
In this figure, the white contours correspond to the {\it uv}-range constrained 1.3 mm continuum.
To indicate the original 1.3 mm high-resolution continuum distribution, we overlaid the 6$\sigma$ contour of  the full {\it uv}-range 1.3 mm high-resolution continuum image in Fig.~\ref{1.3mm_910klambda} (red contour).
To give close-in views of the continuum objects, we divided the whole region into seven subclumps (clump A to G are named CA, CB, CC, CD, CE, CF, and CG), and the subplots are also presented in Fig.~\ref{1.3mm_910klambda}.

In this work, we assumed an identification threshold of 12$\sigma$\footnote{12$\sigma$ is the most negative lobe found in the {\it uv}-restricted image, and hence we do not trust any structure below this threshold as they could be related to artifacts or residual sidelobes.} in the {\it uv}-range constrained continuum image. 
If the 12$\sigma$ contour corresponds to an isolated source and presents no substructure, this was counted as one single source. 
On the contrary, if the 12$\sigma$ contour contains further substructure, we considered different sources within this 12$\sigma$ contour if they have at least one different closed contour. 
In Fig. \ref{dendrogram} we illustrate this identification criterion by indicating the 12$\sigma$ contour for isolated sources along with each first closed contour (above the threshold) that unambiguously separates each identified source\footnote{Since all identified sources are separated by at least one beam (see Fig. \ref{dendrogram}), the identified sources can be considered as observationally distinct objects.}.
Based on this criteria, a total of 33 continuum sources were identified within the surveyed region, and the positions of the identified continuum sources are marked as red crosses in Fig. \ref{1.3mm_910klambda}.
A two-dimensional Gaussian fitting was performed toward each identified source, and the fitting results are listed in Table~\ref{frag_table1}.

Generally, the identified sources are embedded within a more extended environment. 
However, there are some isolated and point-like sources, which seem to be already dynamically decoupled from their parental filamentary structures, and they are probably objects with a different nature.
Therefore, in order to avoid over-identification, we made a classification for all identified continuum sources based on the following two criteria.
First, the red contour in Fig. \ref{1.3mm_910klambda} is 6$\sigma$ of the 1.3 mm continuum with full {\it uv} range; if a source is enclosed by a red contour individually, and the continuum source does not have an extended structure, it is considered as an isolated source, which means that the source was already decoupled from its parental structure. 
Second, if a source has a ratio of intensity to flux density higher than 0.5, it is considered as an unresolved point-like source. 
The isolated and point-like sources were not identified as fragments and were excluded from further analysis and marked with ``$\dagger$'' in Table \ref{frag_table1}. 
Finally, 29 out of 33 continuum sources were identified as fragments.
It should be noted that the terminology of ``fragment'' in this work just stands for the smallest compact continuum sources which follow the identification criteria described above.

%\clearpage

Assuming that the 1.3 mm continuum emission of the identified sources is optically thin, their masses can be estimated using the flux densities as:
\begin{equation}\label{eq1}
M_{\rm source} = \frac{S_{\nu} D^{2}}{B(T_{\rm d}) \kappa_{\nu}},
\end{equation}
where, D = 5128($\pm$1867) pc is the distance to W51\,IRS2 \citep{Reid2019} and $B(T_{\rm d})$ is the Planck function at temperature $T_{\rm d}$. \citet{Ginsburg2017} derived a gas temperature range of 200 K < $T_{\rm gas}$ < 390 K for W51 North by using methanol lines.
In this case, since our surveyed region is much larger than W51 North, we assume a dust temperature of $T_{\rm d}$ = $T_{\rm gas}$ = 200 K and that the dust  temperature is uniformly distributed in the surveyed region (this assumed temperature has also been used in \citet{Goddi2020}). We note that
$\kappa_{\nu}$ is the dust opacity, and the value of $\kappa_{\nu}$ from column 6 of Table 1 in \citet{Ossenkopf1994} was adopted to estimate the
densities of continuum sources ($\kappa$ = 0.008991 cm$^{2}$ g$^{-1}$ at 1.3 mm). Furthermore,
$S_{\nu}$ is the total flux density measured in the 1.3 mm {\it uv}-range constrained continuum image, and it was derived from the Gaussian fitting.
The masses of the continuum sources are listed in Table~\ref{frag_table1}.

As mentioned in Section \ref{spectral index}, the majority of sources detected in the surveyed region have spectral indices of $\sim$2, suggesting that they may be partly optically thick.
\citet{Ginsburg2017} also reported that the central region of W51 North seems to be optically thick.
Assuming that these sources are slightly optically thick, the column density of these sources can be estimated by using $N_{\rm H_{2}}$=$\frac{1}{\kappa_{\nu}}$ = 2.6$\times$10$^{25}$ cm$^{-2}$ at 1.3 mm. The masses of the optically thick sources were estimated by $M_{\rm t}$ = $\pi R^{2} N_{\rm H_{2}}m_{\rm H} \mu$, where $R$ is the radius of the sources derived from the Gaussian fitting, $\mu$ = 2.8 is the mean molecular weight (Kauffmann2008), and $m_{\rm H}$ is the mass of H atom. The masses based on the optically thick assumption are also listed in the last column of Table \ref{frag_table1}.
The typical (median) mass estimated by equation \ref{eq1} is 0.4($\pm$0.3) M$_{\odot}$, and the typical mass estimated with the optically thick assumption is 0.5($\pm$0.3) M$_{\odot}$.
This indicates that the optically thin assumption does not strongly affect our results. 
Hereafter, our calculations are based on the assumption that the dust emission is optically thin.

\section{Analysis}
\label{analyses}
\subsection{Typical separations between fragments}
\label{typical separations between fragments}

\begin{figure}[!h]
        \centering
        \includegraphics[width=9cm]{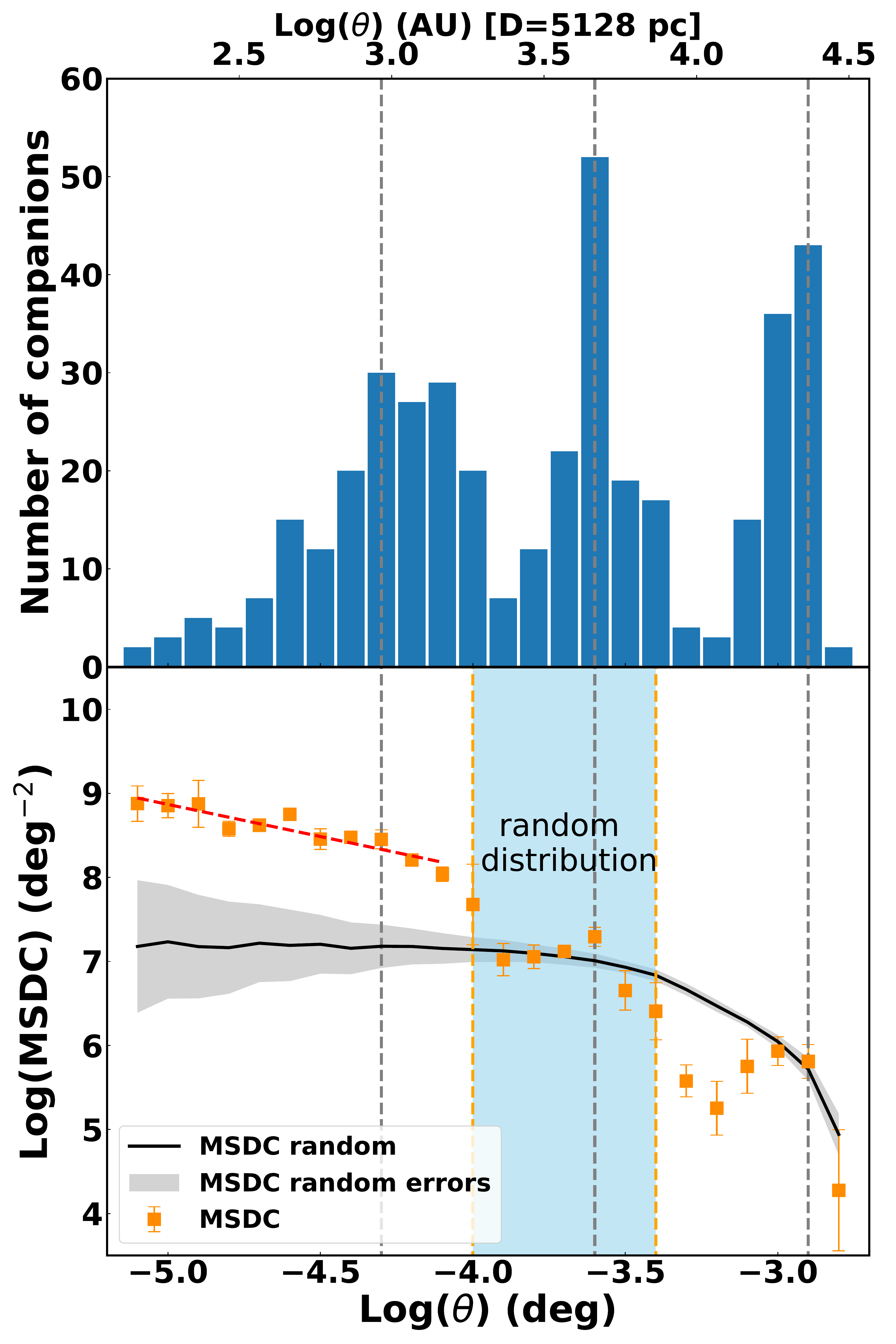}
        \caption{\emph{Upper panel}: Histogram of angular separations of companions. The Y-axis is the number of companions, and the X-axis is angular separations. The gray vertical dashed lines represent the peak number of the companions.
                \emph{Bottom panel}: Orange squares with error bars represent the mean surface density of companions (MSDC) versus the angular
                separation of the identified fragments. The orange vertical dashed lines denote the breaks of the MSDC. 
                The black solid line and the gray region represent MSDC and errors that were derived
                from the 1000 rounds of uniformly, and randomly distributed 29 fragments in the same region. The linear fitting toward the data points at separations smaller than the first break, marked as a red dashed line in the bottom panel, results in a slope of  --0.8($\pm$0.1).}
        \label{MSDC}
\end{figure}

To study the fragmentation process in W51\,IRS2 more quantitatively, we estimated the angular separations of each fragment to every other fragment by using their coordinates listed in Table~\ref{frag_table1}, and we considered a pair of fragments as one companion.
Thus, the 29 identified fragments generate 406 companions, and every companion has its own corresponding angular separation.
We measured the angular separations ($\theta$) of all companions (fragments pairs) in degrees, then we grouped the angular separations of all companions into different annuli with logarithmic bins of log($\theta$) = 0.1, and we plotted the histogram of these grouped separations in the upper panel of Fig.~\ref{MSDC}.
In this figure, one can see that there are three peaks located in the plot with steps in log($\theta$) of --4.3, --3.6, and --2.9, corresponding to 925 AU, 4635 AU, and 23229 AU, respectively.
These three peaks are indicated by gray vertical dashed lines in the upper panel of Fig.~\ref{MSDC}.

A large-scale clump of about $\sim$0.1 pc with its initial physical conditions seems to fragment into medium-scale clumps of $\sim$0.02 pc, then these medium-scale clumps continue to fragment into small-scale objects of $\lesssim$ 0.004 pc and large densities.  
Therefore, the whole large-scale clump presents a hierarchical structure with fragmentation taking place at different scale levels. 
This is called hierarchical fragmentation, and it has been found in different star-forming regions \cite[e.g.][]{Wang2011, Takahashi2013, Zhang2015, Teixeira2016, Kainulainen2017}.
A peak  in the histogram indicates that the number of companions is highest at the corresponding scale, meaning that the clustering phenomenon is present at this scale.

The three peaks of the histogram indicate that there seems to be a multilevel hierarchical structure in our surveyed region.
However, the third peak is meaningless for our further analysis because it corresponds to the largest spatial scale of the whole surveyed area.
Therefore, we only take the first and the second peaks of the histogram into consideration.
To a certain extent, the typical separation corresponding to each peak could also reflect the physical process controlling fragmentation at the corresponding scale. 
For instance, if fragmentation is consistent with a thermal Jeans process, the typical separation defined by the histogram peak should be comparable with its Jeans length.
We compare the typical separations with Jeans lengths in the next section.

The correlation between the mean surface density of companions (MSDC) and angular separations ($\theta$) are generally used
to study the source distribution and fragmentation of molecular clouds \citep{Gomez1993,Larson1995,Bate1998,Kraus2008,Tafalla2015,Palau2018}.
We calculated the MSDC ($\Sigma$($\theta$)) of the identified 29 fragments by following the description of \citet{Simon1997}:
\begin{equation}
\Sigma(\theta) = \frac{N_{p}}{2 \pi r_{\theta} \delta(\theta) N_{f}} \,,
\end{equation}
where $N_{p}$ is the number of fragment pairs (or the number of companions) that fall in the annulus with a radius $r_{\theta}$ and width $\delta(\theta)$, and $N_{f}$ is the total number of fragments.
The logarithmic relationship between $\Sigma(\theta)$ and the angular separation ($\theta$) is shown in the bottom panel of Fig.~\ref{MSDC}.
The orange squares denote the averaged $\Sigma(\theta)$  of each annulus, and the error bars correspond to the standard deviation.
In order to compare the MSDC of our identified fragments with the MSDC corresponding to uniformly, randomly distributed sources,
we ran 1000 rounds of calculations of MSDC for a uniformly random distribution of sources within the same area.
We plotted the averaged result and errors over 1000 rounds of random distributions as a black solid line and gray shaded region in the bottom panel of Fig.~\ref{MSDC}, respectively.

Only the region from --4.0 $\le$ log($\theta$)$\le$ --3.4 shows a flat behavior (within errors), consistent with our observed fragments being randomly distributed (blue area in Fig. \ref{MSDC}). We define a break as the spatial scale where the MSDC is no longer consistent with the random distribution, and it is related to the spatial scale of the parental structure undergoing fragmentation. From the bottom panel of Fig. 3, two breaks are found at log($\theta$) = –4.0 (1845 AU) and log($\theta$) = –3.4 (7346 AU). In the bottom panel of Fig. 3, the orange vertical dashed lines indicate the breaks of $\Sigma(\theta)$.

The values of the MSDC for separations larger than the second break at log($\theta$) = --3.4 (7346 AU) are below the gray area. 
This is because the surveyed area is undergoing hierarchical fragmentation, but our random sources were generated within the entire surveyed region. 
If there was only a one-level fragmentation process, the flat region would extend up to large scales, and only one break would appear in the plot. 
On the other hand, the values of the MSDC for separations smaller than the first break at log($\theta$) = --4.0 (1845 AU) are clearly located well above the gray shaded area, revealing an enhancement of mean surface density, and thus a significant clustering phenomenon. 
In addition, these points at separations smaller than log($\theta$) = --4.0 (1845 AU) present a linear behavior, mainly contributed to by the fragments in W51 North. 
We performed a linear fit toward these points and found a slope of --0.8($\pm$0.1) (the result is plotted as a red dashed line in Fig. \ref{MSDC}). 
This value is very consistent with the values obtained for the linear regime of the MSDC in the Pipe Nebula (--0.68) and the Lupus I cloud (--0.72)  \citep{Roman2019}, where this linear regime is found at similar or slightly larger spatial scales of 2000 -- 8000 AU.

\subsection{Ensembles defined as groups of fragments}
\label{core}

\begin{figure*}
        \centering
        \includegraphics[width=18.4cm,clip]{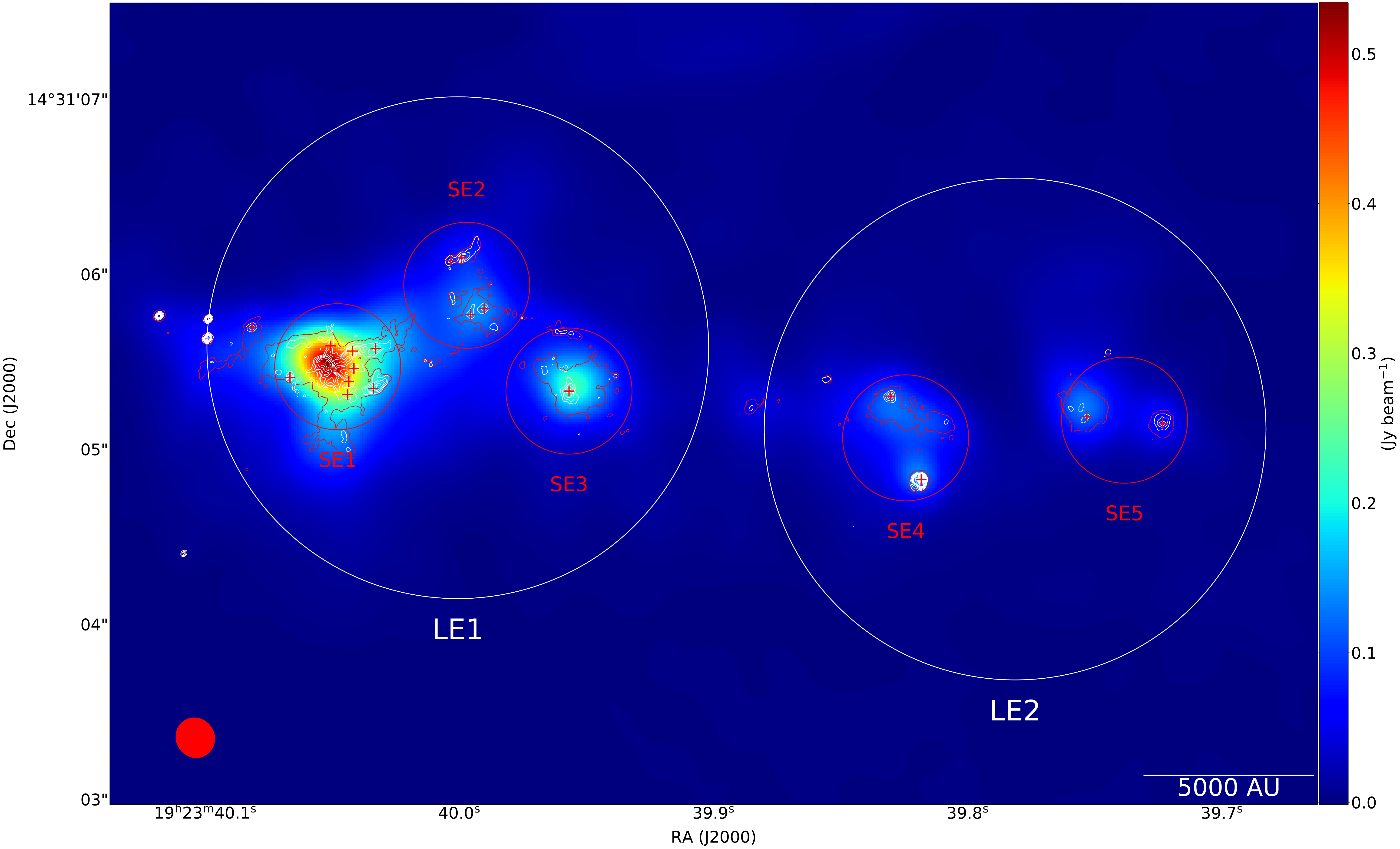}
        \caption{Color image corresponds to the 1.3 mm low-resolution continuum image with the $uv$ range 9--910 k$\lambda$, and white contours indicate the 1.3 mm continuum with a {\it uv} range larger than 910 k$\lambda$. The white contour levels are [-12, -6, 6, 12, 12, 18, 24, 30, 36, 40, 60, 80, 100, 120,140,160]$\times\sigma$ ($\sigma$ = 0.11 mJy beam$^{-1}$), and the beam size is shown in the bottom left corner. The red contour in the plot is the 1.3 mm continuum with a full {\it uv} range, and the contour level is 6$\sigma$ ($\sigma$ = 0.15 mJ beam$^{-1}$). The small and larger circles denote the 
%        {\bf \LEt{ EDITOR: When I first edited this,\ I hadn't realized you used the term SE later on for something else, so I've added the quotation marks back for SE.\ My apologies for the inconvenience.\ However I do suggest that every time you are referring to "small-scale ensembles" that you write "SEs" with quotation marks for clarifiaction purposes.\ Please carry this out on your end throughout your paper.}
%ANSWER FROM AUTHORS: we think that including quotation marks every time we mention SEs or LEs can distract the reader and we strongly prefer to avoid quotation marks once these terms have been introduced.}
"small-scale ensembles" (SEs) and "large-scale ensembles" (LEs), respectively. These SEs and LEs are defined by the spatial scales corresponding to the first and second breaks from the MSDC analysis (see Section~\ref{core} for details). Overall, five SEs and two LEs are defined in the region, and their names are labeled in the figure. The information about these ensembles is listed in Table~\ref{core_table}.}
        \label{fig_core}
\end{figure*}

In Section~\ref{typical separations between fragments}, two  MSDC breaks were found corresponding to 1845 and 7346 AU.
As we mentioned before, these breaks reflect the beginnings of clustering phenomena, and they are related to the size of parental structures that undergo fragmentation.
Thus, we use such spatial scales to define groups of fragments at different spatial scales.       
In this way, we draw a circle with a  radius corresponding to the identified breaks for every fragment.
If a circle includes the center of any other fragment, we consider them to be related fragments that
belong to the same ``parental structure''.
For the convenience of the subsequent discussion, we define an ``ensemble'' as a group of subobjects belonging to the same parental structure (following \citet{Zhang2021} nomenclature).

Based on the above criteria, a total of five ensembles were identified with a spatial scale of 1845 AU.
The center of each ensemble was calculated by taking the geometric average of the positions of the fragments that
belong to the same ensemble, and a spatial scale of 1845 AU was adopted as its radius.
Taking a spatial scale of 7346 AU as a radius, only two ensembles were identified.
Since the spatial scale of the first break is smaller than the second one, we call the ensembles identified using the 1845 AU radius "small-scale ensembles" (SEs). 
Each small ensemble is referred to  as an SE plus numbers.
The ensembles identified using the 7346 AU radius are called  "large-scale ensembles" (LEs).
The coordinate of each LE is the geometric average of the positions of the SEs that belong to the same LE.
The details regarding the ensembles are listed in Table~\ref{core_table}.
The identified ensembles are shown as small and large circles in Fig.~\ref{fig_core}, and the names of SE and LE are labeled in the figure.
It should be noted that every fragment is assigned to a SE (except a fragment in ``CA4'' outside the circle of SE1, which is still  assigned to SE1).

From Fig \ref{fig_core}, the hierarchical fragmentation scenario in this region is well illustrated. 
The entire region fragments into two separated subregions, LE1 and LE2, then each large-scale ensemble fragments into different small-scale ensembles, respectively. 
Finally, each small-scale ensemble splits up into several fragments that have been identified using the {\it uv}-range constrained continuum image.

\subsection{Fragmentation}
\label{fragmentation}

In order to investigate the fragmentation process occurring in our surveyed region, 
we estimated the fragmentation levels and densities of each ensemble.
We defined the fragmentation level of SEs as the number of identified fragments belonging to that SE.
For LEs, we defined their fragmentation level as the number of SEs belonging to the corresponding LE.
The fragmentation level of each ensemble is listed in Table~\ref{core_table}.

Assuming that the identified ensembles are spheres with a uniform density distribution, then their densities can be estimated as follows:
\begin{equation}\label{eq3}
n_{\rm ens} = \frac{M_{\rm ens}}{\frac{4}{3} \pi R^{3} \mu m_{\rm H}},
\end{equation}
where $R$ is the radius of the ensemble, which has been considered to be 1845 AU for SEs and 7346 AU for LEs,
$\mu$ = 2.8 is the mean molecular weight \citep{Kauffmann2008}, $m_{\rm H}$ is the mass of a H atom, and
$M_{\rm ens}$ is the mass of an ensemble, which can be estimated from the total flux density within the circle region in Fig. \ref{fig_core}
(radii of 1845 AU for SEs and 7346 AU for LEs) following equation \ref{eq1}.
A dust temperature of $T$ = 200 K was used to estimate the masses of the ensembles, assuming that the dust
temperature is constantly distributed in the region.
Since the 1.3 mm {\it uv}-constrained high-resolution data are employed to identify fragments and as the 3 mm data are affected by the free-free emission, the data set at 1.3 mm with low angular resolution (0.227$^{\prime\prime}$, Section \ref{1.3 low}) was used to estimate the masses of the ensembles.
This data set has the largest recoverable angular scale of $\sim10^{\prime\prime}$, which covers the surveyed region very well.
As we mentioned before, the high-resolution 1.3 mm continuum image with a {\it uv} range larger than 910 k$\lambda$ was adopted to identify fragments.
In order to estimate the densities of the ensembles from the 1.3 mm low-resolution data, we generated a 1.3 mm low-resolution continuum image using only the {\it uv} range of 9.164 $\sim$ 910 k$\lambda$.
Since the 1.3 mm low-resolution data has a resolution of 0.227$^{\prime\prime}$ (1163 AU), which is smaller than the diameter of SEs,
we directly used the flux density from 1.3 mm low-resolution data to estimate the masses and densities of ensembles.
The estimated densities of SEs and LEs are listed in Table~\ref{core_table}.

As can be seen from Table \ref{core_table}, SE1 has the highest fragmentation level and is also the densest ensemble. 
Other SEs have densities smaller than SE1, and also lower fragmentation levels.
While this is the first result consistent with thermal Jeans fragmentation, further tests need to be performed to make this result more robust.

\begin{table*}
        \caption{\label{core_table}Properties of the identified ensembles of fragments}
        \centering
        \tiny
        \begin{tabular}{lcccccccccccc}
                \hline\hline
                Name$^{a}$ &RA(J2000)&Dec(J2000)       &frag level$^{b}$     &$M_{\rm ens}^{c}$    &$n_{\rm ens}^{c}$     &$\lambda_{J}^{d}$  &M$_{J}^{d}$     &L$_{typical}^{e}$ &M$_{typical}^{e}$         \\
                                           &                        &                                &                                       &$M_{\odot}$                                         &(cm$^{-3}$)                                         &(AU)                  &M$_{\odot}$                      & (AU)                          &$M_{\odot}$         \\
                \hline
                \hline
            \multicolumn{5}{c}{Small-scale ensembles identified by the first break (radius: 1845 AU )}              & &   & & \\
\hline
SE1            &19:23:40.04798 &+14:31:05.4793  &20(5)  &75(55)  &4(1)E+08 &817(112) &0.9(0.1)   &1063(434) &0.7(0.5) \\
SE2            &19:23:39.99720 &+14:31:05.9444  &4(2)   &26(19)  &1.2(0.5)E+08 &1492(310) &1.6(0.3)   &1650(56)  &0.3(0.1) \\
SE3            &19:23:39.95698 &+14:31:05.3403  &1(1)   &30(22)  &1.5(0.5)E+08 &1335(222) &1.5(0.2)   &-- --     &1.1(0.9) \\
SE4            &19:23:39.82453 &+14:31:05.0730  &2(1)   &25(18)  &1.2(0.4)E+08 &1492(249) &1.6(0.3)   &2616(942) &2(1) \\
SE5            &19:23:39.73840 &+14:31:05.1741  &2(1)   &15(11)  &7(3)E+07 &1954(419) &2.2(0.5)   &2322(836) &0.4(0.3) \\
\hline
\multicolumn{5}{c}{Large-scale ensembles identified by the second break (radius: 7346 AU)}     &  &&\\
\hline
LE1           &19:23:40.00072 &+14:31:05.5880  &3(2)  &226(165)  &1.7(0.6)E+07 &3964(700)  &4.4(0.8)    &4574(198)  &30(4)\\
LE2           &19:23:39.78147 &+14:31:05.1235  &2(1)  &80(58)    &6(2)E+06     &6672(1112) &7(1)    &6641(2390) &20(8)\\
\hline
        \end{tabular}
        \tablefoot{
                \\$^{a}$: The ensembles identified by the first break in Fig.~\ref{MSDC}-bottom are called small-scale ensembles (SEs), and the ensembles
                identified by the second break are called large-scale ensembles (LEs) (see Section \ref{core}).
                \\$^{b}$: For SEs, the fragmentation level is defined as the number of fragments associated with each SE. 
                For LEs, the fragmentation level is defined as the number of SEs associated with each LE.
                The uncertainty in the fragmentation level is assumed to be the Poisson error ($\sqrt{N}$).
                \\$^{c}$: The mass and (H$_2$) density were derived from the 1.3 mm low-resolution continuum data with a {\it uv} range of 9 $\sim$ 910 k$\lambda$,
                and they are consequently independent of the 1.3 mm high-resolution data that we used to identify fragments (with a {\it uv} range $\textgreater$ 910 k$\lambda$).
            \\$^{d}$: The Jeans lengths and masses were derived from the densities of SEs and LEs, estimated assuming a temperature of 200 K.
            \\$^{e}$: L$_{typical}$ is the median separation between every two subobjects within each ensemble, and M$_{typical}$  is the median mass of subobjects within each ensemble. The errors correspond to the median absolute deviations. For SE4, SE5, and LE2, there are only two subobjects within each ensemble, therefore, there are no median absolute deviations for them, and their errors were only estimated from the uncertainty of the distance to {W51\,IRS2} (5128$\pm$1867 pc).}
\end{table*}

A second test for thermal Jeans fragmentation is the comparison between the typical fragment separations and masses with the expected values from a Jeans analysis. 
To do this, the densities derived from the 1.3 mm low-resolution data with a $uv$ range of 9--910 k$\lambda$ were used to estimate the Jeans lengths and Jeans masses for all the identified ensembles. 
Assuming an idealized sphere with uniform distributions of density and dust temperature, the Jeans length and Jeans mass can be estimated as follows:
\begin{equation}\label{eq4}
\lambda_{J} = c_{s}\left(\frac{\pi}{G\rho}\right)^{\frac{1}{2}},
\end{equation}
\begin{equation}\label{eq5}
M_{J}=\frac{4}{3}\pi\rho\left(\frac{\lambda_{J}}{2}\right)^{3},
\end{equation}
where $c_{s}$ is the sound speed at temperature $T$ = 200 K, $G$ is the gravitational constant, and $\rho$ is the mass density. The estimated Jeans lengths and Jeans masses for the ensembles are listed in Table \ref{core_table}.
We also estimated the typical (median) separation between every two subobjects within each SE and LE, along with the typical (median) masses of the subobjects, and listed the results in Table \ref{core_table}.

From Table \ref{core_table}, SE1 results in a $\lambda_{J}$ = 817($\pm$112) AU and $M_{J}$ = 0.9($\pm$0.1)  M$_{\odot}$. 
The Jeans length obtained for SE1 is consistent (within errors) with the typical separations of fragments within SE1.
Additionally, as we mentioned before, the first peak shown in the upper panel of Fig.~\ref{MSDC} is mainly contributed to by SE1 (W51 North), and the scale identified by this first peak (925 AU) is also consistent with its Jeans length.
Similarly, a comparison of the Jeans lengths of all the other identified SEs and LEs to the typical observed separations indicate that they are fully consistent within errors. 

Regarding the typical masses of the fragments in SEs, these are generally similar to or even smaller than the Jeans mass. In no case is the difference (after taking the uncertainties reported in the table into account) larger than a factor of {\bf 3}, and this could be explained by a combination of both dust temperature and dust opacity uncertainties. Thus, for SEs it is plausible to think that the typical measured masses are also consistent with the Jeans mass. 
As for LEs, the masses of the subobjects (SEs) are in general larger by up to a factor of 5 (LE1) or 2 (LE2) than the Jeans mass. Although the typical mass of subobjects for LE1 is clearly higher than the corresponding Jeans mass, by only considering a temperature of 300 K instead of 200 K in the Jeans analysis, the difference would only be a factor of 3 (instead of 5).
These results are further discussed in Section~\ref{discussion1}.

\section{Discussion}\label{discussion}

\subsection{Thermal Jeans fragmentation could be at work in W51\,IRS2}\label{discussion1}

In Section~\ref{fragmentation} two tests for the thermal Jeans fragmentation scenario were performed. First, it was studied if the fragmentation level of the ensembles was related to the density of these ensembles. From this study, it was found that SE1, the ensemble associated with W51\,North, presented both the highest fragmentation level along with the highest density. In addition, the Jeans length and mass estimated for SE1 are very comparable to the measured median separations and masses of the fragments in SE1. %Thus, from our previous analysis it seems that SE1 is undergoing fragmentation according to a thermal Jeans process.
Thus, from our previous analysis it seems that the observed fragmentation in SE1 is consistent with thermal Jeans fragmentation.
        
While this scenario seems quite clear for SE1, the low fragmentation levels in the other SEs is more intriguing. In fact, one can estimate the number of fragments expected for each SE by calculating the Jeans number:
        
\begin{equation}
N_\mathrm{Jeans} = \frac{M_\mathrm{ens}\,\mathrm{CFE}}{M_\mathrm{Jeans}},
\end{equation}
        
where CFE stands for core formation efficiency and is a factor accounting for the fact that, given a density structure, it is not expected that it is entirely fragmented at the same time, but that the highest density substructures fragment first and thus there must always be a fraction of the parental structure which has not fragmented yet  \citep[e.g.,][]{Palau2015,Vazquez-Semadeni2019}.
%
%{\bf The number of observed fragments in SE1 is 10, which was assumed to be $N_{\rm Jeans}$. The Jeans mass ($M_{\rm Jeans}$) and total mass ($M_{\rm ens}$) of SE1 can be found in Table \ref{core_table}.}
%If we calculate the Jeans number for SE1 it is found that $N_\mathrm{Jeans} = 83$, while the number of observed fragments is $\sim20$. 
If we calculate the Jeans number for SE1 assuming that CFE = 100\%, then $N_{\rm Jeans}$ = 83, while the number of observed fragments is 20, indicating that assuming CFE = 100\% is probably not correct. In order to have $N_{\rm Jeans}$ = 20 in SE1,
the CFE must be $\sim$24\%, which is consistent with the values found by \citet{Palau2015} for a sample of massive dense cores. Thus, one can calculate  $N_\mathrm{Jeans}$ for the other SEs, adopting the same CFE obtained for SE1. The resulting $N_\mathrm{Jeans}$ are 4, 5, 4, and 2 for SE2, SE3, SE4, and SE5, respectively. This expected number of fragments for the SEs matches the observed number of fragments well, especially for the case of SE2 and SE5, while the case deviating more strongly from this prediction is SE3 where only one fragment is identified while five fragments were expected.

This suggests that either other mechanisms in addition to thermal pressure are at work in SE3, or that the Jeans mass has been underestimated for SE3. In the first case, one could think of both turbulence or magnetic fields as possible additional supporting mechanisms against gravity. However, the role of turbulence is greatly debated in the community because in order to act as an effective form of support against gravity, it should be isotropic, which is highly questioned because turbulence is characterized by an energy spectrum containing most of the energy at large scales \citep[see Section 2.2 of][see also Dobbs et al. 2005, Burkhart \& Mocz 2019]{Vazquez-Semadeni2019}. In addition, this has not been supported by observational tests (e.g., Palau et al. 2015). Regarding the support offered by the magnetic field, there are a number of theoretical and numerical works suggesting that the magnetic field should be efficient at suppressing fragmentation \citep[e.g.,][]{Boss2004,Vazquez-Semadeni2005,Commercon2011a,Hennebelle2011}, 
%also: Ziegler 2005; Price & Bate 2007; Peters et al. 2011; Bailey & Basu 2012; Myers et al. 2013; Boss & Keiser 2013, 2014; Girichidis et al. 2018; Hennebelle & Inutsuka 2019
and also very recent observational works seem to be consistent with this prediction \citep[e.g.,][]{AnezLopez2020,Palau2021}. However, for the case of W51\,IRS2, there are magnetic field measurements by \citet{Koch2018} and the magnetic field does not seem to be particularly strong in SE3 compared to SE1.
%SE3 does not seem to present a particularly dominating role of\LEt{ I suggest replacing "of" by "on" or "in" if possible.} the magnetic field compared to SE1. 

In conclusion, it seems that the most plausible explanation for the low fragmentation found in SE3 is that it might be warmer than the other SEs naturally inhibiting fragmentation. If a temperature of 400 K is assumed in the Jeans calculation, the Jeans mass increases to $\sim4$~M$_{\odot}$, and the Jeans number decreases to $\sim2$, consistent with our observations within the Poisson uncertainty. 
In order to test if SE3 is actually hotter than the other ensembles of W51 IRS2,
we checked the 1.3 mm molecular line data of W51 IRS2 (high-resolution dataset), and vibrational transitions of CH$_{3}$CN (v8=1) have been found in absorption in W51 North, W51 d2, and SE3 (Tang et al., in preparation). This indicates that these three regions should have gas temperatures higher than the other identified ensembles, and this is consistent with the assumption that SE3 has a temperature higher than 200 K.
We note that the high temperatures required in the Jeans analysis to explain our observations indicate that W51\,IRS2 is experiencing a 
%\LEt{ Please avoid the use of slashes. For details, refer to Sect. 2.9 of the language guide.\ Another option could be replacing the slash with "and".}
reduced or suppressed fragmentation event, that is
%\LEt{ e.g./i.e. should be written out in full when part of the main text (not inside parentheses or figure legends). i.e. should be replaced by "that is" or similar when in main text. See the entries for e.g. and i.e. in Section 2.1, "Note 2" of the language guide https://www.aanda.org/for-authors/language-editing/2-main-guidelines.} 
if the ambient or parental cloud had not been heated by nearby massive stars, W51\,IRS2 would most likely undergo a more powerful fragmentation event in terms of the number of fragments.

In summary, if temperatures for the Jeans analysis are adopted to be in the range 200--400 K, then our observations are very consistent with the expected numbers for thermal Jeans fragmentation. Our observations suggest that the fragmentation in W51\,IRS2 is thermally inhibited, as already suggested by \citet{Ginsburg2017}, and that a thermal Jeans fragmentation process is probably at work with an unusually high temperature
%\LEt{ I suggest writing "along with the temperature being unusually" if possible or rephrasing.}
%with the particular case that the temperature is unusually high in the region 
due to the stellar feedback from the nearby massive stars.

\subsection{A cluster-like spatial distribution of fragments in W51\,North (SE1)}\label{discussion2}

In Section~\ref{typical separations between fragments}, a linear regime in the MSDC plot of W51\,IRS2 was reported for separations $<1845$~AU (Fig.~\ref{MSDC}-bottom), and the fragments corresponding to these separations are essentially the fragments associated with W51\,North. The value of the slope of the linear regime is related to the physical process determining the separation between sources. As already explained in \citet{Larson1995} and \citet{Simon1997}, a particular scale is identified in several star-forming clouds above which a slope of $-0.6$ was measured in the MSDC plot, associated with the 
%\LEt{ US convention uses double quotation marks to indicate a special use of a word or phrase the first time it is used in the text, but thereafter quotations marks are not needed. If the special meaning is otherwise clear, or indicated by 'so-called' or similar, the quotation marks are not needed.}
"clustering regime." On the other hand, below this particular scale, a slope of $-2$ was measured for the same clouds, associated with the so-called "binary regime". The clustering regime is associated with the fragmentation of the cloud, while the binary regime is affected by additional processes such as the dynamical interactions between the different members of a multiple system.  
The value of the slope found here for W51\,North (SE1), --0.8$\pm$0.1 (Section~\ref{typical separations between fragments}), is fully consistent with the expected value for the clustering regime in other clouds of the Milky Way such as Taurus, Ophiucus, the Pipe Nebula, Lupus I, or OMC-1S \citep{Gomez1993,Simon1997,Palau2018,Roman2019} and clearly shallower than the typical value of the binary regime. 
Although W51 North is very young and part of the fragments identified here could merge with time, the fact that the slope of the MSDC plot for W51 North precisely corresponds to the expected value for the clustering regime suggests that a protocluster is in the making in W51 North. However, one should keep in mind that if a fraction of the fragments merge with time, the slope of the linear regime could change. Further kinematic studies should be carried out to explore the possible evolution of the fragments.

It is important to compare the slope and scale of the linear regime found in Fig. \ref{MSDC} for W51 North to other star-forming regions also presenting the clustering regime.
The scale at which the MSDC changes from a clustering to binary regime (what we call here the "clustering-to-binary scale") is found to vary from region to region and, interestingly, \citet{Simon1997} reported an anticorrelation between the clustering-to-binary scale and the stellar density of each cloud. In order to test if both W51\,IRS2 and OMC-1S fit the anticorrelation found by \citet{Simon1997}, we made a rough estimate for the stellar density in W51\,IRS2 and OMC-1S by considering that the fragments identified here are true (precursors of) protostellar objects. Although most of the fragments identified in this work have no reported signs of stellar activity such as outflows, such an estimate for the stellar density is probably accurate as a first order of magnitude because of several reasons. 
First, the identified fragments have masses that are generally consistent with the Jeans mass (within uncertainties) and therefore could be gravitationally unstable. 
Second, thanks to the {\it uv} restriction used to identify fragments (Section~\ref{fragments}), only the fragments that are already very compact and centrally peaked have been considered, and we have left out the larger noncentrally peaked structures which could be transient structures. 
%{\bf The density of $10^{6} - 10^{8}$ cm$^{-3}$ is high enough to allow the formation of a high-mass star \citep{Motte2018}.}
Using the data reported in Table~\ref{frag_table1}, we calculated the densities of the fragments and their median value is 8$\times$10$^9$~cm$^{-3}$. 
Such a high density is precisely the density at which the first hydrostatic core, the direct precursor of the protostellar object, formed in the model of \citet{Larson1969}, and in subsequent more recent works \citep[e.g.,][]{Commercon2011b,Bhandare2018}.
Third, although there are no infrared or X-ray counterparts associated with the identified fragments due to a lack of observations with a similar resolution as our 1.3 mm continuum data, \citet{Palau2018} identified fragments in OMC-1S adopting the same method as described in Section \ref{fragments}, and they found that 50\% of them are associated with star-forming signposts.
Therefore, based on the above reasons, it is feasible that an important fraction of the fragments identified here are already protostellar or will become protostellar in the near future. 
Another fraction of the fragments detected here might not collapse either because of thermal support, mergers of fragments, or tidal forces stretching the fragments \citep[e.g.,][]{Lee2018,Hennebelle2019}.

Thus, the stellar density for both W51\,IRS2 and OMC-1S was estimated by dividing the number of identified fragments in each region by the volume corresponding to the area where these fragments have been identified (assuming they are spherically distributed).  The resulting stellar densities for both W51\,IRS2 and OMC-1S, along with the previously reported stellar densities for Taurus, Ophiucus, and the Trapezium, are listed in Table ~\ref{stdens_table}. %where a 50\% of uncertainty is assumed for the stellar density estimate.
Given the reasons listed in the previous paragraph, a 50\% uncertainty is assumed for the stellar density estimation.
        
Regarding the clustering-to-binary scale, for both W51\,IRS2 and OMC-1S, the MSDC clearly reveals a linear regime with a slope in the range of $-0.75$-- $-1$, which is well below the slope for the binary regime, but the transition to the binary regime itself is not detected. Thus, we consider that the lower limit of our range of separations in the MSDC plot is the upper limit of the clustering-to-binary scale for W51\,IRS2 and OMC-1S. 
        
        \begin{figure}
                \centering
                \includegraphics[width=9cm,clip]{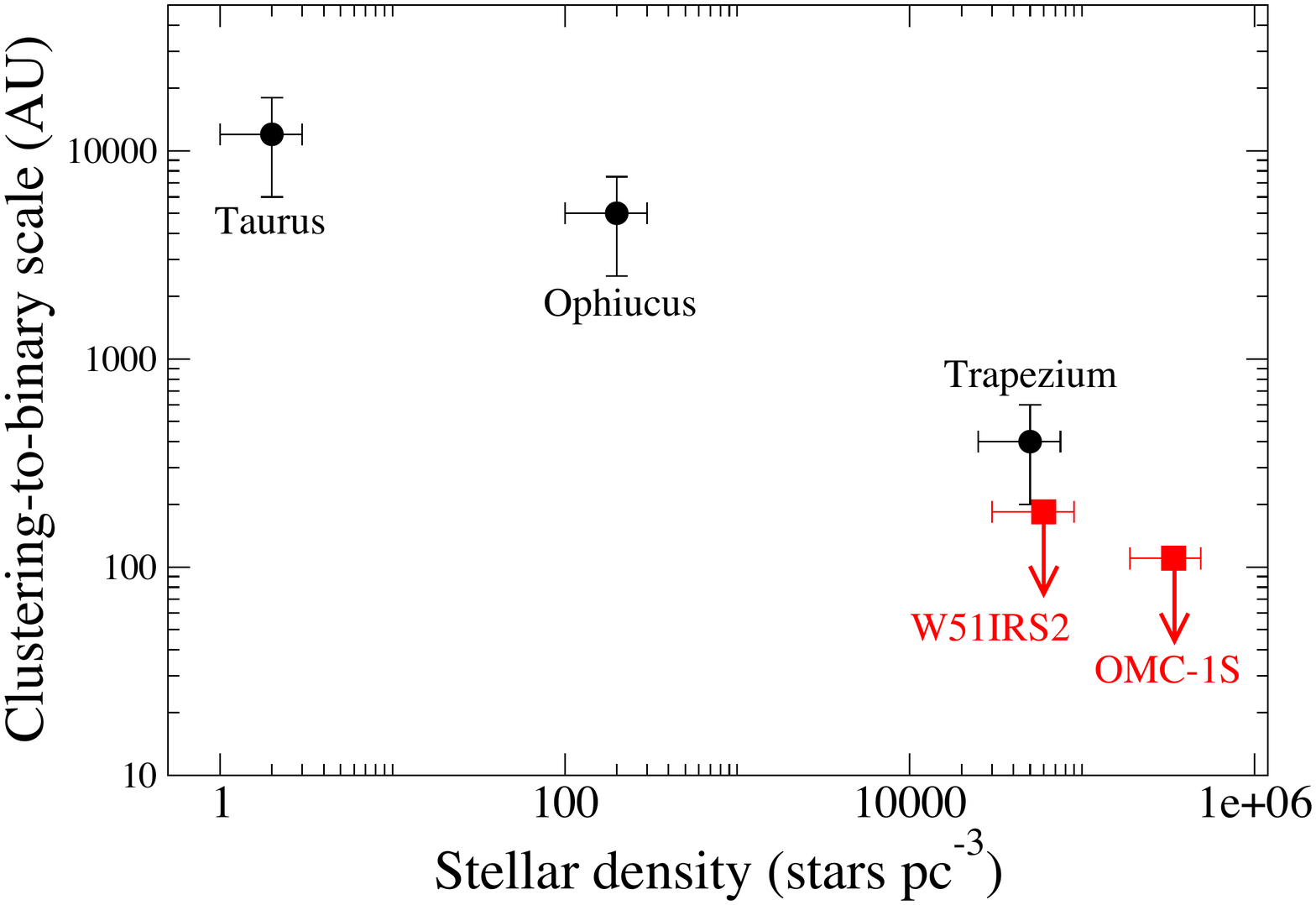}
                \caption{Anticorrelation between the clustering-to-binary scale in the MSDC plot and stellar density for different clouds.
                        The data point of W51\,IRS2 is taken from this work; the data point of OMC-1S is taken from \citet{Palau2018}; and the data points of Taurus, Ophiucus, and Trapezium are taken from \citet{Gomez1993} and \citet{Simon1997}.}
                \label{fig_correlation}
        \end{figure}
        
The anticorrelation discovered by \citet{Simon1997} is presented in Fig.~\ref{fig_correlation}, where the data for both W51\,IRS2 and OMC-1S have already been included. The figure shows that the two regions added here follow the anticorrelation  reasonably well between the clustering-to-binary scale and the stellar density, but now at higher stellar densities. 
%In particular, the W51\,IRS2 point is especially close to the point corresponding to the Trapezium, suggesting that W51\,IRS2, and more specifically W51\,North, could become such a kind of cluster in the future. 
This finding, along with the slope value for the linear regime associated with W51\,North, which is very similar to the values reported for the clustering regime in other clouds of the Galaxy, indicates that W51\,North seems to be forming a protocluster with spatial properties very similar to the properties of other clustered regions in the Galaxy. 
        
In Section~\ref{discussion1}, it is shown that fragmentation in W51\,IRS2 is consistent with a thermal Jeans process, but with the particular case of high temperatures (200--400 K), probably produced by the stellar feedback from nearby massive stars. Therefore, fragmentation in W51\,IRS2 has probably been thermally inhibited (as already suggested by \citet{Ginsburg2017}), and much more fragments might have formed in the region if it had not been previously heated to such high temperatures. However, the inhibited fragmentation does not seem to prevent the formation of a protocluster associated with W51North as the spatial distribution of the associated fragments (most likely being or becoming protostellar) are fully consistent with the spatial distribution of typical cluster-like regions in the Galaxy.

Finally, we would like to mention that not only does W51\,IRS2 fit the anticorrelation, but OMC-1S also corresponds to stellar densities that are even higher. The fact that the slopes of the clustering regime in the MSDC plot are very similar in all the regions compiled here (between $-0.6$ and $-1$), which in turn cover a very broad range of stellar densities, is consistent with the fractal dimension (ratio of the change in detail to the change in scale) of the molecular cloud structure being similar for a wide range of gas densities and spatial scales, as already suggested by \citet{Simon1997}.

\begin{table}
        \caption{\label{stdens_table}{Stellar densities and clustering-binary scales for clouds in the literature and this work}}
        \centering
        \tiny
        \begin{tabular}{lcccc}
                \hline\hline
                &Mean           &Stellar                &Clustering-to-binary     \\
                &radius         &density                        &scale in MSDC                    \\
                Region     &(pc)                        &(stars pc$^{-3}$)      &(AU)                                 &Refs.$^\mathrm{a}$   \\
                \hline
                Taurus                  &0.85           &2.0                    &12000                                 &G93   \\
                Ophiucus                        &$-$            &200                    &5000                                 &S97  \\
                Trapezium               &$-$            &$5\times10^4$  &400                                 &S97  \\
                W51\,IRS2               &0.05           &$6\times10^4$  &$<180$                         &This work   \\
                OMC-1S                  &0.03           &$34\times10^4$          &$<110$                         &P18  \\
                \hline
        \end{tabular}
        \tablefoot{
                \\{$^\mathrm{a}$ G93: G\'omez et al. (1993); S97: Simon (1997); P18: Palau et al. (2018)}.
        }
\end{table}

\section{Conclusions}
\label{conclusions}
To study the fragmentation properties of the W51\,IRS2 region, ALMA data from three projects have been used.
Based on our analyses, the main findings of this work are summarized as follows:

   \begin{enumerate}
      \item A spectral index map was derived from 3 mm and 1.3 mm high-resolution continuum images.
      Based on the spectral index map, we found that most areas in the surveyed region have spectral indices around 2.0,
      with only two regions, W51 d2 and CC1, associated with spectral indices around 0.2 indicative of free-free emission,
      indicating that the majority of the detected sources might be partly optically thick.

      \item A 1.3 mm $\it uv$-range constrained high-resolution continuum image ({\it uv} range $>$ 910 k$\lambda$) was generated to identify fragments,
     % \LEt{ I suggest writing "contain" or using the following format for clarification purposes "; the + noun + only contains...".}
%     which only contains continuum emission from the structures with scales smaller than 0.1$^{\prime\prime}$ ($\sim$ 500 AU).
     which is sensitive only to structures with angular scales smaller than 0.1$^{\prime\prime}$ ($\sim$ 500 AU).
      Based on a 12$\sigma$ identification threshold, a total of 33 spatially separated continuum objects have been identified and 29 of them are defined as fragments, with masses ranging from 0.2 to 2 M$_{\odot}$.

      \item  The mean surface density of companions (MSDC) analysis reveals two spatial scales at 1845 AU and 7346 AU associated with overdensities, suggesting a two-level hierarchical fragmentation. We used these two spatial scales to define ensembles as groups of fragments. Five small-scale ensembles (SEs, corresponding to a radius of 1845 AU) and two large-scale ensembles (LEs, corresponding to a radius of 7346 AU) are defined in our surveyed region.

      \item The fragmentation level and densities of both SEs and LEs have been estimated. 
        The ensemble with the highest density, associated with W51\,North, corresponds to the ensemble with the highest fragmentation level. 
      Combining this with a Jeans analysis, it seems that fragmentation in SEs and LEs is consistent with a thermal Jeans process if high temperatures in the range from 200--400 K are considered, naturally explaining the small number of fragments along with their masses and separations.

      \item A linear regime is found in the MSDC plot for separations $<1845$~AU, with a slope of --0.8$\pm$0.1. Such a slope is very similar to the slope observed in previous works for the clustering regime in the MSDC, and the anticorrelation found by \citet{Simon1997} between the transition scale from the clustering to the binary regime and the stellar density seems to hold for the higher stellar densities studied here.
   \end{enumerate}

Overall, W51\,IRS2 seems to be a case of thermal Jeans fragmentation operating at high temperatures due to stellar feedback from nearby massive stars, implying a thermal inhibition of fragmentation in the region. However, such an inhibition does not seem to be powerful enough to prevent the formation of a protocluster in W51\,North, for which the spatial distribution of the future protostars seems to be fully consistent with the spatial distribution of well-known clusters in the Galaxy.

\begin{acknowledgements}
        The authors deeply appreciate the crucial comments from the referee, which have undoubtedly improved the quality of this paper.
        A.P. is grateful to Enrique V\'azquez-Semadeni for very insightful discussions.
        This paper makes use of the following ALMA data: project IDs: 2013.1.00994.S, 2015.1.01596.S and 2017.1.00293.S.
        ALMA is a partnership of ESO (representing its member states), NSF (USA) and NINS (Japan), together with NRC (Canada), NSC and
        ASIAA (Taiwan), and KASI (Republic of Korea), in cooperation with the Republic of Chile. The Joint ALMA Observatory is operated by ESO, AUI/NRAO, and NAOJ.
        L.A.Z. acknowledges financial support from CONACyT-280775,  and  UNAM-PAPIIT  IN110618  grants,  M\'exico.
        A.P. acknowledges financial support from the UNAM-PAPIIT IN111421 grant, the Sistema Nacional de Investigadores of CONACyT, and from the CONACyT project number 86372 of the `Ciencia de Frontera 2019’ program, entitled `Citlalc\'oatl: A multiscale study at the new frontier of the formation and early evolution of stars and planetary systems’, M\'exico.
        Sheng-Li, Qin is supported by the key research project of the National Natural Science Foundation of China  (project number: 12033005).
        Mengyao Tang is supported by China Scholarship Council (CSC), and PhD research startup foundation of Chuxiong Normal University.
\end{acknowledgements}


%\end{document}

\begin{appendix} %First appendix
\section{ALMA simulations}
\label{AppendixA}
In Section \ref{fragments}, we used the 1.3 mm high-resolution {\it uv}-range constrained continuum image to identify the continuum objects.
However, cutting off short baselines affects the {\it uv} coverage and may create artifacts in the final continuum image. 
In order to study whether cutting-off baselines shorter than 910 k$\lambda$ can significantly affect our results, we simulated ALMA observations using the ``SIMOBSERVE'' task in the CASA package.

\begin{figure*}[!h]
        \centering
        \includegraphics[width=\hsize]{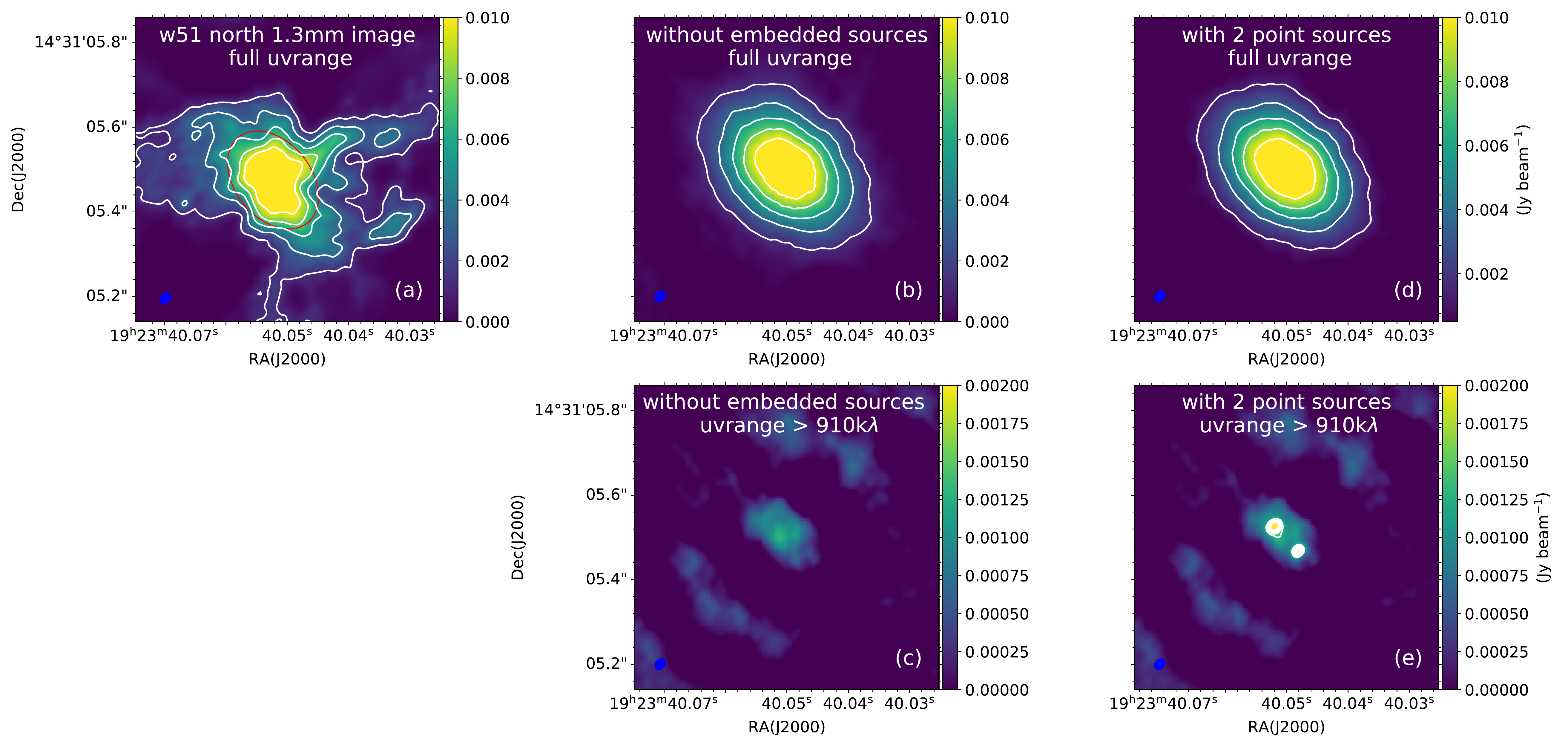}
        \caption{\emph{Panel (a)}: 1.3 mm continuum image of the W51 North region with a  full {\it uv} range. 
                The red ellipse is given by Gaussian fitting toward W51 North. This ellipse has been adopted as a model to perform ALMA simulations. \emph{Panel (b)}: Simulated full {\it uv}-range continuum image without any embedded source. \emph{Panel (c)}: Simulated {\it uv}-range constrained ($\textgreater$ 910 k$\lambda$) continuum image without any embedded sources. \emph{Panel (d)}: Simulated full {\it uv}-range continuum image with two point sources.  \emph{Panel (e)}: Simulated {\it uv}-range constrained ($\textgreater$ 910 k$\lambda$) continuum image with two point sources. The contour levels in all the panels start from 12$\sigma$ (1.32 mJy beam$^{-1}$). The residual map of the 1.3 mm continuum image generated by the TCLEAN task of CASA has been added to the simulated images to realistically simulate the noise.}
        \label{simulation1}
\end{figure*}

However, CASA does not provide the specific antenna list used for the 1.3 mm high-resolution observations. 
%To obtain the simulated data similar to the realistic observations as much as possible, 
Thus, we manually generated an antenna list completely consistent with the 1.3 mm high-resolution observations to perform these simulations.
Fig \ref{simulation1} presents the results of two different simulations with full {\it uv} coverage and constrained {\it uv} coverage. 
Panel (a) shows the 1.3 mm high-resolution continuum image with a full {\it uv} range.
Firstly, we performed a Gaussian fitting to W51 North, then we took the ellipse obtained by Gaussian fitting as the source size to simulate an elongated structure with a comparable spatial scale to W51 North.
Secondly, we ran two different simulations.
In the first run, the model did not include any embedded compact objects. 
In the second run,  two point sources were included, and the typical intensity of the identified continuum sources in Table \ref{frag_table1} were considered to be the peak intensity of the point sources.
Finally, the TCLEAN task in CASA was employed to generate continuum images for the two simulations.
The realistic residual map from TCLEAN of the 1.3 mm continuum observations was added to the simulated continuum images as the noises.
Panels (b) and (d) show continuum images with a full {\it uv} range generated from the two different simulations, respectively.
According to panels (b) and (d), there is almost no difference between the two images.
The continuum images of two different simulations with only a {\it uv} range $\textgreater$ 910 k$\lambda$ are presented in panels (c) and (e), respectively.
One can see that two compact sources appear in panel (e), but no compact source is found in panel (c), indicating that cutting off baselines shorter than 910 k$\lambda$ does not create strong artifacts in the 1.3 mm high-resolution image, but this allows us to identify the embedded sources.

\begin{figure*}[!h]
        \centering
        \includegraphics[width=1.0\hsize]{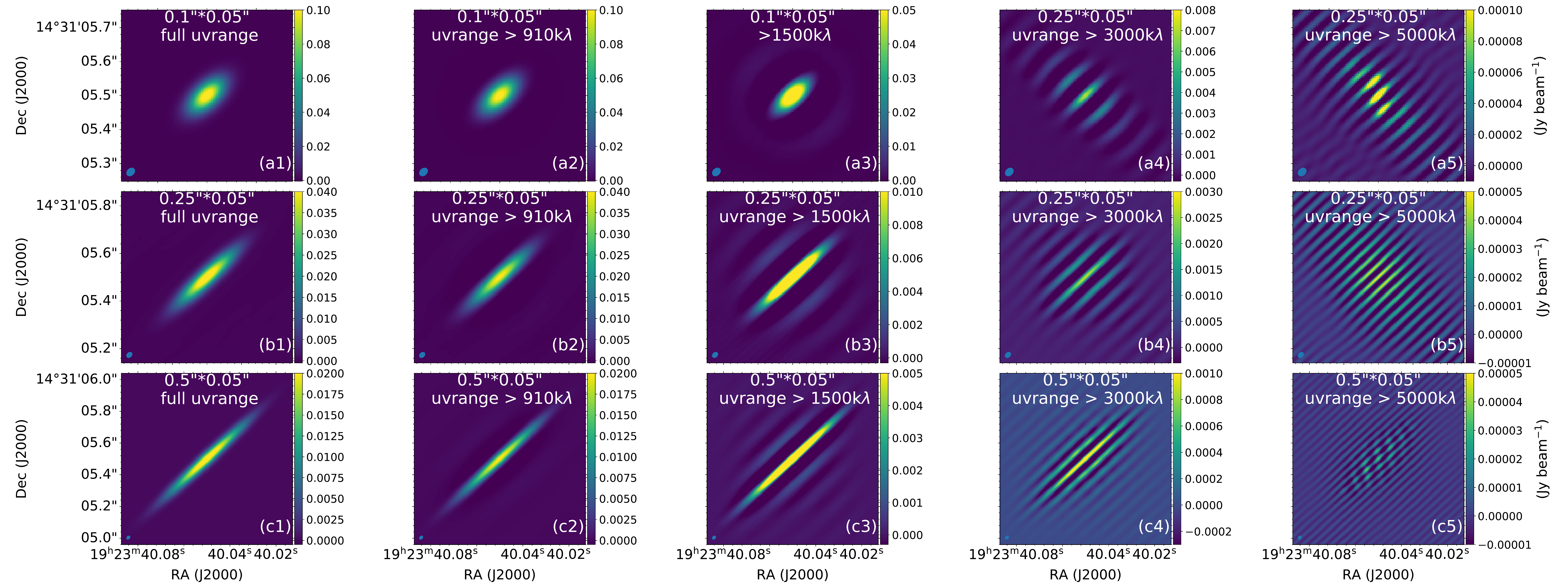}
        \caption{Simulated continuum images with different source sizes and {\it uv} ranges. No simulated sources have embedded point sources. The source sizes and {\it uv-}range cutoff thresholds are labeled in each panel. The beam size is shown in the bottom left corner of each panel.}
        \label{simulation3}
\end{figure*}

Short baselines are sensitive to large-scale emission, and cutting off short baselines may lead to severe missing flux problems. 
To test how our continuum images change with the {\it uv-}range cutoff, we performed several simulations toward different elongated structures with a variety of {\it uv-}range cutoffs, as shown in Fig \ref{simulation3}.
In panels (a1) - (a5) of Fig. \ref{simulation3}, the source size has been set as 0.1$^{\prime\prime}$$\times$0.05$^{\prime\prime}$ (ratio of the major axis to minor axis of 2:1), which is similar to the typical size (0.06$^{\prime\prime}$$\times$0.03$^{\prime\prime}$) of continuum objects identified in Section \ref{fragments}. 
This simulation can be used to investigate how the morphology of a typical compact source changes with a {\it uv-}range cutoff.
In panels (b1) - (b5), the source size has been set to 0.25$^{\prime\prime}$$\times$0.05$^{\prime\prime}$ (ratio of the major axis to minor axis of 5:1), which reveals how an elongated structure changes with a {\it uv-}range cutoff.
In panels (c1) - (c5), the source size has been set to 0.5$^{\prime\prime}$$\times$0.05$^{\prime\prime}$ (ratio of the major axis to minor axis of 10:1), which presents how an extremely elongated structure changes with {\it uv-}range cutoffs.
The first to fifth columns of Fig. \ref{simulation3} present the continuum images with a full {\it uv} range, 910 k$\lambda$, 1500 k$\lambda$, 3000 k$\lambda$, and 5000 k$\lambda$ {\it uv}-range cutoff, respectively.
From panels (a1) and (a2) of Fig. \ref{simulation3}, one can see that the intensity and morphology are almost the same in both a full {\it uv }range and 910 k$\lambda$ cutoff, indicating that a 910 k$\lambda$ {\it uv-}range cutoff does not induce severe missing flux problems for the typical source sizes of our observations.
When comparing panels (b1), (b2), (c1), and (c2), we found that none of the elongated structures are split in the 910 k$\lambda$ {\it uv-}range cutoff, which indicates that cutting off baselines shorter than 910 k$\lambda$ does not create significant artifacts.
Raising up the {\it uv-}range cutoff threshold induces interference fringes, and some artifacts indeed appear when cutting off a {\it uv} range shorter than 5000k$\lambda$ as shown in panel (c5). 
However, such a {\it uv-}range cutoff leads to severe missing flux problems and the peak intensities of artifacts are just $\sim$5$\times$10$^{-4}$ Jy beam$^{-1}$, which is certainly below our 12$\sigma$ (1.32$\times$10$^{-3}$ Jy beam$^{-1}$) identification threshold.
Therefore, we are confident that cutting off the baselines shorter than 910 k$\lambda$ does not create artifacts in our 1.3 mm high-resolution image, and the continuum objects identified in Section \ref{fragments} are most likely real compact {\bf sources}. 

\section{Fragments identified using the dendrogram algorithm}
\label{AppendixB}
\begin{figure*}
        \centering
        \includegraphics[width=\hsize]{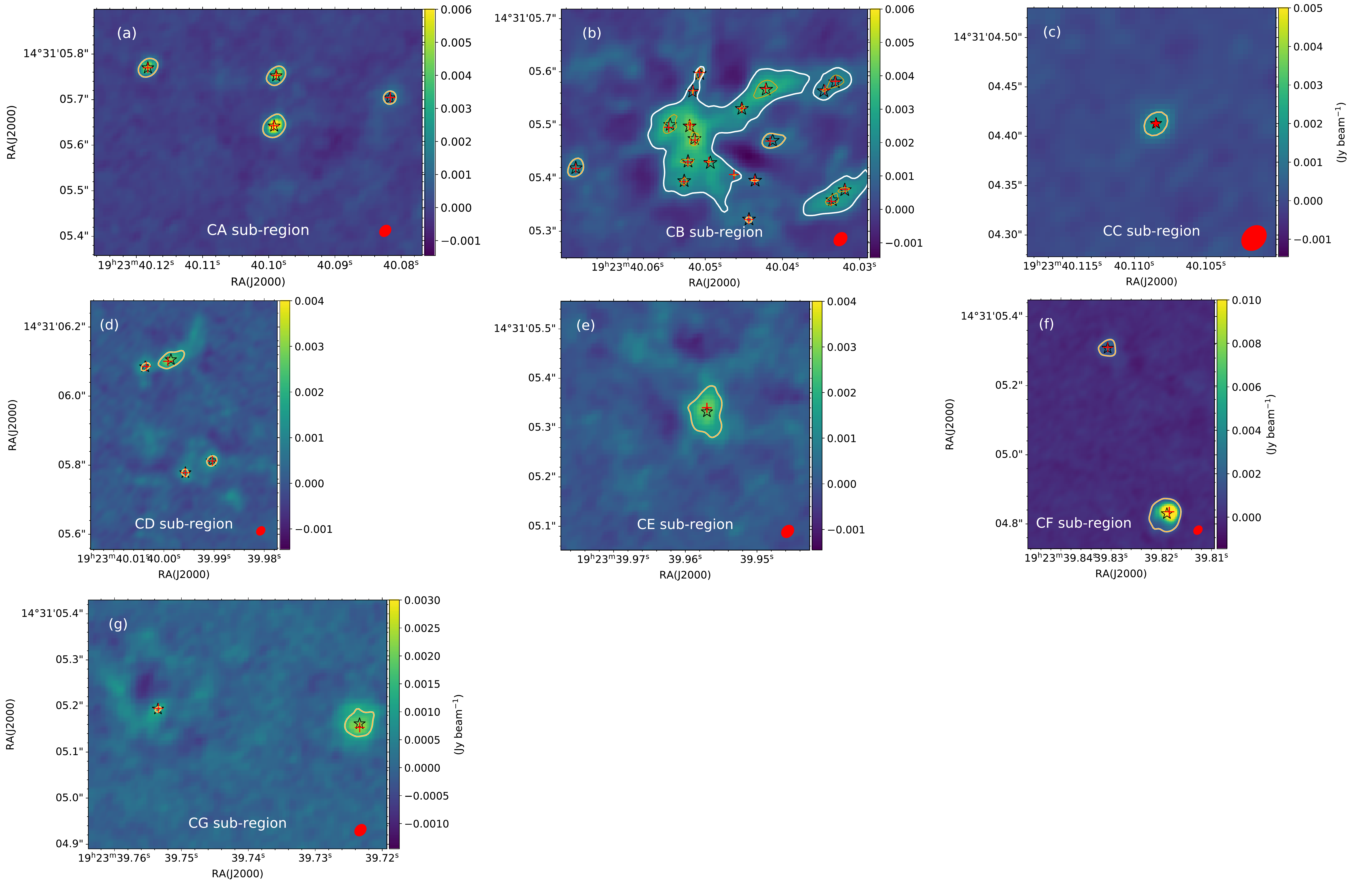}
        \caption{Panels (a) to (g) correspond to subregions CA to CG as in Fig \ref{1.3mm_910klambda}. 
                The continuum objects identified in this work (Section 3.2) are marked with red crosses, and the continuum objects identified by the dendrogram algorithm are marked with black stars in each panel. The color scale is the 1.3 mm {\it uv}-range constrained image, and the white contour corresponds to the 12$\sigma$ (1.32 mJy beam$^{-1}$) identification threshold. The beam size is shown in the bottom right corner of each panel. A total of 32 continuum objects have been identified by the dendrogram algorithm in the surveyed region. The first closed contour above the threshold that separates each identified source is marked as an orange contour to illustrate the ``closed contour'' identification method described in Section \ref{fragments}.}
        \label{dendrogram}
\end{figure*}

The continuum sources listed in Table \ref{frag_table1} were identified through an intensity threshold and a "closed contour" criterion applied to a {\it uv}-restricted image.
In each plot of Fig. \ref{dendrogram}, the first closed contour above the threshold that separates each identified source is also presented as an orange contour.
To explore whether this identification method can affect the results of our work, we adopted the python-based continuum objects identification algorithm of ``dendrogram\footnote{http://www.dendrograms.org/}'' to identify fragments in the surveyed region. 
The identification threshold was set to 12$\sigma$ (1.32 mJy beam$^{-1}$), which is the same as the threshold used in Section \ref{fragments}. 
The differencing threshold (min\_delta parameter) of the dendrogram was set to 0.1$\sigma$ (0.011 mJy beam$^{-1}$).
We considered the ``leaves'' given by the dendrogram algorithm as the continuum sources.
A total of 32 continuum sources were identified by the dendrogram algorithm, and we plotted them in Fig. \ref{dendrogram}. 
The sources identified in Section \ref{fragments} are also shown in Fig. \ref{dendrogram}. 
We found that the majority of continuum sources identified by the dendrogram algorithm are indeed consistent with the sources identified in Section \ref{fragments}, and only one source (CB10) in subregion CB has no counterpart from the dendrogram algorithm.
This indicates that our identification method used in Section \ref{fragments} yields results comparable to other identification methods.
There is a little offset between the positions of the continuum objects defined by the Gaussian fitting and the dendrogram method. 
This offset may be caused by the fact that the dendrogram algorithm adopts the continuum peak as the central position of the identified sources, while in Section 3, we used a Gaussian fitting to obtain the central positions of the continuum objects.
Taking the beam size into consideration, these offsets do not affect our results or discussion in Sections 3, 4, \& 5.
%{\bf In order to illustrate our ``closed contour'' identification method described in Section \ref{fragments}, we plotted the first closed contour as blue contour in each panel.}

\end{appendix}
\end{document}